\documentclass[apj]{emulateapj}
\usepackage[sort&compress]{natbib}
\usepackage{bm}
\usepackage{mathrsfs}
\usepackage{mathtools}
\usepackage{txfonts}
\usepackage{graphicx}
\usepackage{rotating}
\usepackage{xcolor}
\usepackage[lined]{algorithm2e}

\newcommand\T{\rule{0pt}{2.6ex}}       
\newcommand\B{\rule[-1.2ex]{0pt}{0pt}} 

\begin{document}
 
\title{Improving Orbit Estimates for Incomplete Orbits with a New Approach to Priors -- with Applications from Black Holes to Planets}

\author{K. Kosmo O'Neil$^1$, G. D. Martinez$^1$, A. Hees$^{1,2}$, A. M. Ghez$^1$, T. Do$^1$, G. Witzel$^1$,  Q. Konopacky$^3$, E.E. Becklin$^1$, D. S. Chu$^1$, J. R. Lu$^4$, K. Matthews$^5$, S. Sakai$^1$} 
\affil{$^1$Department of Physics and Astronomy, University of California, Los Angeles, Los Angeles, CA 90095, USA \\
$^2${SYRTE, Observatoire de Paris, Universit\'e PSL, CNRS, Sorbonne Universit\'e, LNE, 61 avenue de l'Observatoire 75014 Paris, France}\\
$^3$Center for Astrophysics and Space Sciences, University of California, San Diego, La Jolla, CA 92093, USA\\
$^4$Department of Astronomy, University of California, Berkeley, Berkeley, CA 94720, USA \\ 
$^5$Division of Physics, Mathematics, and Astronomy, California Institute of Technology, Pasadena, CA 91125, USA}

\begin{abstract}
\noindent We propose a new approach to Bayesian prior probability distributions (priors) that can improve orbital solutions for low-phase-coverage orbits, where data cover less than $\sim$40\% of an orbit. In instances of low phase coverage -- such as with stellar orbits in the Galactic center or with directly-imaged exoplanets -- data have low constraining power and thus priors can bias parameter estimates and produce under-estimated confidence intervals. Uniform priors, which are commonly assumed in orbit fitting, are notorious for this. We propose a new observable-based prior paradigm that is based on uniformity in observables. We compare performance of this observable-based prior and of commonly assumed uniform priors using Galactic center and directly-imaged exoplanet (HR 8799) data. The observable-based prior can reduce biases in model parameters by a factor of two and helps avoid under-estimation of confidence intervals for simulations with less than $\sim$40\% phase coverage. Above this threshold, orbital solutions for objects with sufficient phase coverage such as S0-2, a short-period star at the Galactic center with full phase coverage, are consistent with previously published results. Below this threshold, the observable-based prior limits prior influence in regions of prior dominance and increases data influence. Using the observable-based prior, HR 8799 orbital analyses favor low eccentricity orbits and provide stronger evidence that the four planets have a consistent inclination of $\sim$30$^{\mathrm{o}}$ to within 1-$\sigma$. This analysis also allows for the possibility of coplanarity. We present metrics to quantify improvements in orbital estimates with different priors so that observable-based prior frameworks can be tested and implemented for other low-phase-coverage orbits.
\\

\noindent\textit{Subject headings}: Galaxy: center -- Galaxy: fundamental parameters -- methods: statistical -- planets and satellites: fundamental parameters

\end{abstract}


\section{Introduction}\label{sec:intro}
The advent of adaptive optics (AO) has enabled orbital parameter estimation for individual objects in several new and exciting astrophysical systems such as the Galactic nuclear star cluster (NSC) and directly-imaged exoplanet systems. At the center of the Milky Way, the orbit of S0-2 (period = 16 years) has provided the best evidence to date for the existence of supermassive black holes and has now begun to constrain alternative theories of gravity \citep{Ghez2000, Ghez2003, Ghez2005, Ghez2008, Eckart2002, Schoedel2002, Schoedel2003, Eisenhauer2003,   Gillessen2009b, Gillessen2017, Genzel2010, Meyer2012,  Boehle2016,  Grould2017, Hees2017, Parsa2017}. Over time, it has become possible to measure stellar orbits at larger Galactocentric radii and thereby study the dynamical structure of the Galactic NSC -- the only such system for which orbital studies of central black holes and their host galaxies are possible \citep[e.g.][]{Ghez2005, Gillessen2017}. Similarly, AO has opened up a new window in the field of exoplanets. Orbital motions of exoplanets imaged with AO have enabled the first studies of the 3-D dynamical structures of exoplanet systems, which can give insights into the formation and evolution of giant planets \citep[e.g.][]{Marois2008, Marois2010, Chauvin2012, Currie2011, Currie2012a, Esposito2013, Maire2015, Pueyo2015, Rameau2016, Konopacky2016, Zurlo2016, Wertz2017}. However, the majority of orbital measurements at the Galactic center (GC) and of directly-imaged exoplanets have incomplete orbital phase coverage.
 
It is difficult to infer accurate orbital estimates for objects with low orbital phase coverage. In these cases, Bayesian prior probability distributions (priors) can easily dominate parameter estimates. In orbit analyses, uniform priors in model parameters are commonly assumed \citep[e.g.][]{Ghez2008, Boehle2016, Gillessen2017}; however, uniform priors are subjective as they depend on a choice of the parameterization of the model. Other prior forms that still depend on model parameters have also been used, including isotropic orientations as suggested by \citet{Ford2006} and uniformity in the Thiele-Innes elements \citep{Lucy2014}. However, \citet{Lucy2014} showed that when data are not rigorously constraining ({\em e.g.} when observations sweep out an angle that is less than $\sim$40\% of the projected 360$^{\mathrm{o}}$ orbit), such subjective model-based priors can lead to biases in parameter estimates and can produce inaccurate confidence intervals.

We aim to develop a prior that is less subjective, and thus more objective, than what has previously been used in orbit modeling. While the concept of using an objective prior has not been deeply explored in orbit modeling, there is a rich literature in statistics in which objective-prior frameworks have been developed. Although priors cannot be truly objective since any probability distribution contains {\em some} information, objective-prior frameworks aim to minimize prior influence and thus maximize data influence on parameter inference. The Jeffreys prior \citep{Jeffreys46}, based on the Fisher information content, is one such paradigm that is designed to be invariant under parameterization. However, in multiple dimensions, the Jeffreys prior has been shown to be inconsistent with other objective prior frameworks such as reference analysis or location-scale invariance \citep{Bernardo2005}.   For a prior to be considered objective, it must inherently sample relevant regions of parameter space \citep{Jeffreys1946, NeymanScott48, Hartigan64, Jaynes68, Bernardo2005, Berger2009, Pierini2011}. In other words, it requires consistent sampling of regions of parameter space favored by the data, which ideally would ensure unbiased parameter estimates and accurate confidence intervals. Reference analysis \citep{Berger2009} accomplishes this by maximizing the relative expected information between posterior and prior distributions in the asymptotic limit of large data sets. However, most real-life experiments only produce a subset of this asymptotically large dataset and thus can have properties that differ from those of the larger dataset. Consequently, priors based on this asymptotic assumption ({\em e.g.} those from reference analysis) can lead to unintended statistical consequences such as misleading inferences or biases \citep{Gelman2017}.  In addition, such analyses are difficult to calculate, except in simple examples \citep{Pierini2011}. 

In this paper, we propose a new approach to priors that can improve orbital estimates for low-phase-coverage orbits. We refer to these new priors as {\em observable-based priors} as they are based on uniformity in the observables rather than in model parameters. This prior requires all measurements to be equally likely before observation, which promotes consistent sampling and thus limits prior influence in regions of prior dominance and increases data influence. 

We present the general definition of our observable-based prior in Sect. \ref{sec:prior_generalform}, and describe the specific form of the prior for orbit modeling applications in Sect. \ref{sec:prior_orbit}.  We describe the observational data used for this analysis in Sect. \ref{sec:obsData}, and use simulations based on these data to test the performance of the observable-based prior compared to that of standard uniform priors in Sect. \ref{sec:testing}. We then present updated orbital analyses of S0-2 and the HR 8799 planets in Sect. \ref{sec:application}, and discuss the scientific impact in Sect. \ref{sec:discussion}.
	
\section{Observable-based Priors}\label{sec:prior}   

\subsection{General Form}\label{sec:prior_generalform}
We construct a prior that is as objective as possible to promote unbiased parameter estimates and accurate confidence intervals, but that is not based on a hypothetically large data set like in reference analysis \citep{Berger2009}. Defining priors with respect to quantities that can be physically measured ensures a priori that the underlying assumptions (likelihood and priors) make physical sense and that possible observations are not biased before observing data. In other words, a prior should neither de-emphasize areas of parameter space that could be observed, nor emphasize areas of parameter space that cannot be observed. There should theoretically be an equal probability of obtaining observations in all regions of parameter space that are possible to observe, which motivates the construction of a prior based on uniformity in the observables rather than on model parameters.

Observable-based priors assume a prior in observable space, $\mathcal{P}(\mathscr{O})$, which can be transformed into a prior in model-parameter space, $\mathcal{P}(\mathscr{M})$, by inverting the integral
\begin{equation}
\label{eq:obsprdef}
\mathcal{P}(\mathscr{O}) = \int d \mathscr{M} \delta (\mathscr{O} - f_{\mathscr{O}}(\mathscr{M})) \mathcal{P}(\mathscr{M})
\end{equation}
to solve for $\mathcal{P}(\mathscr{M})$. Here, $\mathscr{O} = f_{\mathscr{O}}(\mathscr{M})$ is the definition of the set of observables and $\mathscr{M}$ is the set of model parameters. Observable-based priors require that external  prior knowledge and knowledge of the experimental design are encapsulated in $\mathcal{P}(\mathscr{O})$ rather than in $\mathcal{P}(\mathscr{M})$. In this paper, we assume a one-to-one transformation for simplicity. For problems where $\mathscr{O}$ and $\mathscr{M}$ have the same dimensionality, inverting Equation \ref{eq:obsprdef} gives:
\begin{equation}
  \mathcal{P}(\mathscr{M})  =  \mathcal{P}(\mathscr{O}) \left|\frac{\partial(\mathscr{O})}{\partial(\mathscr{M})}\right| \label{eq:modelprior}.
\end{equation}
Here, ${\partial(\mathscr{O})}/{\partial(\mathscr{M})}$ is the Jacobian defined by the transformation from observable space to model parameter space.  If the dimensionality of $\mathscr{M}$ is larger than that of $\mathscr{O}$, then extra constraints must be introduced to produce an unique prior.  Alternatively, if the dimensionality of $\mathscr{O}$ is larger, then the observables themselves are not independent of each other.  In general, measurements are not independent and identically distributed for orbit analyses, and thus the number of observables increases with the number of measurements. Each observable depends on the epoch of observation, making the distributions $\mathcal{P}(\mathscr{O})$ and $\mathcal{P}(\mathscr{M})$ dependent on the epoch $t$ ({\em e.g.} $\mathcal{P}(\mathscr{O} | t)$ and $\mathcal{P}(\mathscr{M} | t)$).  This dependence makes it impossible to uniquely specify $\mathcal{P}(\mathscr{O} | t)$ and $\mathcal{P}(\mathscr{M} | t)$ at every observed epoch. Instead, we marginalize over a PDF of the observing schedule, $\mathscr{P}(t)$, to specify a marginal distribution $\mathcal{P}(\mathscr{M})$. This marginal distribution can be approximated as a sum over the epochs of observation:
\begin{eqnarray}
  \mathcal{P}(\mathscr{M}) & = & \int \mathcal{P}(\mathscr{M} | t) P(t) dt \\
			    & \approx & \sum_j \mathcal{P}(\mathscr{M} | t_j)\label{eq:obprsum}.
\end{eqnarray}

We define our prior such that the probability of a set of observables $\mathscr{O}$ at time $t$, $\mathcal{P}(\mathscr{O} | t)$, is uniform within a range that is proportional to the measurement error. Because the likelihood is invariant under coordinate translation, a flat prior distribution in observable space ensures that the posterior is also invariant under the same assumptions \citep{Jaynes68}. Observable-based priors of other forms in observable coordinates can also be assumed, depending on the form of the likelihood.  Assuming uniformity in observable space, Equation \ref{eq:modelprior} conditioned on time $t$ becomes:
\begin{eqnarray}
  \mathcal{P}(\mathscr{M} | t) & = & \mathcal{P}(\mathscr{O} | t) \left|\frac{\partial(\mathscr{O}(t))}{\partial(\mathscr{M})}\right| \label{eq:obprdt}\\
                           & \propto & \frac{1}{\prod_i \sigma_i(t)} \left|\frac{\partial(\mathscr{O}(t))}{\partial(\mathscr{M})}\right|,\label{eq:obprprod}
\end{eqnarray}
where $\prod_i \sigma_i(t)$ is the product of the measurement uncertainties for the set of observables at time $t$. That the specified form of the prior is inversely proportional to the measurement uncertainties implies that, in practice, the prior assigns different weight to information on observables carried by each measurement. In other words, more emphasis is placed on data points with smaller measurement uncertainties. Note that the above prior is identical to the Jeffreys prior \citep{Jeffreys46} for a single epoch, though this similarity ends for datasets that span multiple epochs.
\footnote{
The Jeffreys prior for multiple epochs is: 
\begin{equation}
 \mathcal{P^{J}} = \sqrt{\left|\sum_i \frac{\partial(\mathscr{O}(t_i))}{\partial(\mathscr{M})} \cdot \Sigma_i^{-1} \cdot \frac{\partial(\mathscr{O}(t_i))}{\partial(\mathscr{M})} \right|} \label{eq:jeffreysprior},
\end{equation}
where $\Sigma$ is the covariance matrix ($\Sigma_{i, j} = \sigma^2_{i, j}$). 
}
Under the above assumptions, it follows from Equations \ref{eq:obprsum} and \ref{eq:obprprod} that the observable-based prior in model-parameter space becomes:
\begin{eqnarray}
  \mathcal{P}(\mathscr{M})  & \approx & \sum_j \mathcal{P}(\mathscr{M} | t_j) \\
			    \label{eq:obs_prior}
                           & \propto & \sum_j \frac{1}{\prod_i \sigma_i(t_j)} \left|\frac{\partial(\mathscr{O}(t_j))}{\partial(\mathscr{M})}\right|.  
\end{eqnarray}

This observable-based prior is based on experimental design, though {\em not} on actual observed data. This idea has precedence in Bayesian statistics, as the Jeffreys prior is similarly based on a form of the likelihood. While our prior relates to the Jeffreys prior for a single epoch, it has different philosophical motivation. Rather than being based on the abstract concept of information content, our observable-based prior is based on the practical idea that there should be an equal probability of obtaining observations in regions of parameter-space that are possible to observe. In addition, there should be no probability of obtaining observations in regions of parameter space that are not possible to observe, thereby avoiding the pitfalls of asymptotic-based priors \citep{Gelman2017}. Thus, the observable-based prior is less subjective than commonly-assumed model-based priors since it does not directly impose constraints on model parameters, but is weakly-informative rather than truly objective in the sense that it imposes constraints on possible observed data.

\subsection{Observable-based Priors in Orbit Modeling}\label{sec:prior_orbit}
We construct an observable-based prior of the general form defined in Sect \ref{sec:prior_generalform} specific to a Keplerian point-potential orbit model in which the central mass, $M_{\mathrm{cent}}$, dominates.  In future work, we will extend the model to include post-Newtonian parameters in General Relativity such as the relativistic redshift. In the Keplerian model described here, seven global parameters that describe the gravitational potential and six Keplerian orbital parameters comprise the model $\mathscr{M}$. The seven global parameters are the central mass ($M_{\mathrm{cent}}$), the line-of-sight distance to the primary ($R_o$), its position on the sky ($x_o$, $y_o$), and its three-dimensional velocity ($v_{x,o}$, $v_{y,o}$, $v_{z,o}$).  The six Keplerian orbital elements, further detailed by \citet{Ghez2005}, are the angle to the ascending node ($\Omega$), argument of periapse ($\omega$), inclination (\textit{i}), orbital period (\textit{P}), time of closest approach ($T_o$), and eccentricity (\textit{e}). The model  $\mathscr{M}$ is thus defined as: 
\begin{equation}\label{eq:model}
\mathscr{M} = \{ M_{\mathrm{cent}},  R_o,  x_o,  y_o,  v_{x,o},  v_{y,o},  v_{z,o},  \Omega,  \omega,  i,  P,  T_o,  e \}.
\end{equation}

Observational data $\mathscr{D}$ constrain the set of model parameters $\mathscr{M}$. For data with only astrometric observations (e.g. directly-imaged planets), the set of measured observables $\mathscr{D}$ at time $t$ is:
\begin{equation}\label{eq:dataPlanets}
\mathscr{D} = \{x(t), y(t)\}, 
\end{equation}
where $x$ and $y$ describe the orbiting body's position on the sky with respect to the position of the primary, $x_o$ and $y_o$. For data with both astrometric and RV observations (e.g. GC data), 
\begin{equation}\label{data}
\mathscr{D} = \{x(t), y(t), v_z(t)\},
\end{equation}
where $v_z$ is the orbiting body's line-of-sight velocity. Note that $x$, $y$, and $v_z$ refer to the position and velocity of the orbiting body, while $x_o$,  $y_o$,  $v_{x,o}$,  $v_{y,o}$, and $v_{z,o}$ describe the position and velocity of the primary (in the GC case, the SMBH). Relative positions are characterized by the angular separation between the primary and the orbiting body, projected onto the plane of the sky. While the position of the primary can correspond to a particular right ascension (RA) and declination (DEC), relative RA and DEC coordinates would depend on an object's location on the curved coordinate system, requiring an additional transformation. In the small angle approximation, there is effectively no curvature in the coordinate system and thus there is approximately a linear relationship between ($\Delta x, \Delta y$) and ($\Delta$RA, $\Delta$DEC). In such cases, RA and DEC or any other linearly-related coordinate system could be used. As long as the transformation between coordinate systems is linear, a uniform distribution in one implies a uniform distribution in the other. In this work, relative astrometric positions, defined as $\mathscr{D} = \{x(t), y(t)\}$, are used since a linear transformation converts measurements from pixel-space to positions on the sky (defined with respect to the position of the primary). Relative astrometry is commonly measured in this way \citep[e.g.][]{Konopacky2016, Gillessen2017}. Further, a functional form exists for $x(t)$ and $y(t)$ such that a prior can easily be formulated.

The {\it measured} observables $\mathscr{D}$ are related linearly to a set of {\it orbital} observables $X(E)$, $Y(E)$, $V_X(E)$, and $V_Y(E)$ that describe the position and velocity in the orbital plane by
\begin{eqnarray}
 x(t) & = & \frac{B}{R_o } X(E) + \frac{G}{R_o} Y(E) + v_{x,o}t +  x_o \label{eq:x} \\ 
 y(t) & = & \frac{A}{R_o } X(E) + \frac{F}{R_o } Y(E) + v_{y,o}t +  y_o\label{eq:y} \\ 
 v_z(t) & = & C V_X(E) + H V_Y(E) + v_{z,o}\label{eq:vz},
\end{eqnarray}
where the orbital observables $X(E)$, $Y(E)$, $V_X(E)$, and $V_Y(E)$ \citep[e.g.][]{Ghez2003} are defined as
\begin{eqnarray}
X & = & a(\cos  E - e) \label{eq:X},\\
Y & = & a(\sqrt{1-e^2} \sin E) \label{eq:Y}, \\
V_X & = & -\frac{\sin E}{1 - e \cos E} \sqrt{\frac{\mbox{G}M}{a}} \label{eq:VX},\\
V_Y & = &\frac{\sqrt{1 - e^2} \cos E}{1 - e \cos E} \sqrt{\frac{\mbox{G}M}{a}} \label{eq:VY},
\end{eqnarray}
and $A$, $B$, $C$, $F$, $G$, and $H$ are the Thiele-Innes constants \citep[e.g][]{Hartkopf1989, Wright2009}, defined by
\begin{eqnarray}
 A & = & +\cos\Omega  \cos\omega - \sin\Omega  \sin\omega  \cos i \\
 B  & = & +\sin\Omega  \cos\omega + \cos\Omega  \sin\omega  \cos i\\
 C  & = &  \sin\omega  \sin i \\
 F  & = & -\cos\Omega  \sin\omega - \sin\Omega  \cos\omega  \cos i\\
 G  & = & -\sin\Omega  \sin\omega + \cos\Omega  \cos\omega  \cos i\\
 H & = &  \cos\omega \sin i.
\end{eqnarray}
The factor of $R_o$ in Equations \ref{eq:x} and \ref{eq:y} is included to convert from angular distance to physical distance. Here, $M$ is the mass of the system (approximated as $M_{\mathrm{cent}}$), G is the gravitational constant, a is the semi-major axis of the orbit, defined as
\begin{equation}
a = \left(\frac{\mbox{G} M P^2 }{4 \pi^2}\right)^{1/3},
\end{equation}
and {\it E} is the eccentric anomaly, defined in terms of the mean anomaly {\it m} as
\begin{equation}
E - e \sin E = \frac{2 \pi}{P} (t - T_o) = m.
\end{equation}

To obtain a form of the observable-based prior in model-parameter space, as defined in Sect. \ref{sec:prior_generalform}, we must transform from observable space to model-parameter space. Since the measured and orbital observables are linearly related, a uniform distribution in one parameter set implies a uniform distribution in the other and thus we can use the orbital observables for the transformation. In this paper, we explore two observable-based priors:  (a) a prior based solely on astrometric observables $\mathcal{O} = \{X, Y\}$ , and (b) a prior based on both astrometric and RV observables $\mathcal{O} = \{V_X, V_Y\}$. 

To perform a one-to-one transformation, we apply the transformation only to a subset of the model parameters, $\mathcal{M} = \{e, P\}$. Applying the transformation to e and P is the natural choice for the reasons described below, though in theory there are innumerable forms that an observable-based prior could take. This one-to-one transformation causes the observable-based prior to be conditioned on the parameters that appear in the Jacobian transformation ($M_{\mathrm{cent}}$ and $T_o$, see Equations \ref{eq:jac1} and \ref{eq:jac2}).

For multiple star or planet fits, the orbital parameters are unique for each orbiting body, while the central potential parameters are shared. Thus, to generalize our prior to multiple star or planet fits, the priors on the global parameters should be specified separately, and we base our transformation only on the orbital parameters. 

For each of the global parameters that act as translation parameters ($x_o$, $y_o$, $v_{x, o}$, $v_{y, o}$, and $v_{z, o}$), we assume a uniform prior -- the standard objective prior for parameters that only shift the probability distribution \citep{Bernardo2005}:
\begin{equation}
 \mathcal{P}(x_o, y_o, v_{x,o}, v_{y,o}, v_{z,o}) \propto \mathrm{constant}.
\end{equation}
For the prior based solely on astrometric observables, we assume a log-uniform prior on $R_o$ to ensure scale invariance since $R_o$ acts as scale parameter. However, we assume a uniform prior on $R_o$ for the prior based on astrometric and RV observables because of the $R_o$ dependence that appears in Equation \ref{eq:obsprior} below. A log-uniform prior is also assumed on $M_{\rm{cent}}$. For (a) the prior based solely on astrometric observables and (b) the prior based on both astrometric and RV observables, these assumptions on $R_o$ and $M_{\rm{cent}}$ yield
\begin{equation}
 \mathcal{P}(R_o, M_{\mathrm{cent}}) \propto 
 \left\{
 \begin{array}{ll}
  \frac{1}{M_{\rm{cent}}} \frac{1}{R_o}  &  \rm{(a)} \\
  \frac{1}{M_{\rm{cent}}}  & \rm{(b)}.
\end{array}
\right .
\end{equation}
Alternatively, without RV measurements, $R_o$ and $v_{z,o}$ cannot be independently constrained, and should be fixed if these values are known. In this paper, for directly-imaged planets where we have astrometric but not RV data, we assume that the mass of the primary is well-defined and that positional measurements are defined relative to the host star. We therefore fix all global parameters that describe the central potential in this case.

Of the remaining six orbital parameters, four can easily be specified separately. $T_o$ is a translation parameter with respect to time, so we again assume a standard objective prior of uniformity:
\begin{equation}\label{eq:Tprior}
 \mathcal{P}(T_o) \propto \mathrm{constant}.
\end{equation}
Because a uniform distribution in $X$ and $Y$ guarantees uniformity in $x(t)$ and $y(t)$ as described above, the $\omega$, $\Omega$, and $i$ prior components can be specified separately since $X$ and $Y$ do not depend on these angular parameters (Equations \ref{eq:X} -- \ref{eq:VY}), reducing the choice of model parameters to $e$ and $P$. For the angular parameters, we assume a prior whose spatial orientation is uniform in direction ({\it e.g.} uniform in cosine of the inclination), as suggested by \citet{Ford2006}:
\begin{equation}
 \mathcal{P}(i) \propto \sin(i),
\end{equation}
and
\begin{equation}
 \mathcal{P}(\omega, \Omega) \propto \mathrm{constant}.
\end{equation}

For the prior based on astrometric observables, we set $\mathcal{O} = \{X, Y\}$ and $\mathcal{M} = \{e, P\}$ in Equation \ref{eq:obs_prior}.  The transformation from $X$ and $Y$ to $P$ and $e$ produces a Jacobian:
\begin{eqnarray}\label{eq:jac1}
\lefteqn{J_{\rm{astro}} \equiv \frac{\partial (X, Y)}{\partial (e, P)} = -\left(\frac{(\mbox{G}M)^2P}{2\pi^4}\right)^{1/3}} \\ 
  & \times & \frac{[2(e^2 - 2) \sin E + e ( 3 m + \sin 2E) + 3m \cos E]}{6\sqrt{1 - e^2}} \nonumber.
\end{eqnarray}
Similarly, if given RV data, the RV observables ($\mathcal{O}=\{V_X, V_Y\}$) produce a Jacobian of:
\begin{eqnarray} \label{eq:jac2}
\lefteqn{J_{\rm{RV}} \equiv \frac{\partial (V_X, V_Y)}{\partial(e, P)} = \frac{(\mbox{G}M\pi)^{2/3}}{3 (2)^{1/3}P^{5/3}}} \\
 &\times&\frac{[\sin E (e^2 \cos 2E + 3 e^2 -2) + \cos E (6m - 2e\sin E)]}{\sqrt{1 - e^2} (1 - e \cos E)^3} \nonumber.
\end{eqnarray}
Finally, the $\prod_i \sigma_i(t_j)$ term in Equation \ref{eq:obs_prior} must be defined within the orbital observable space in units of the measured uncertainties of $x$, $y$, and $v_z$.  Since $\prod_i \sigma_i(t_j)$ represents some volume in parameter space, equivalent volumes between measured and orbital observable space can be related by the Jacobian that defines the transformation between these observable spaces. Since the orbital and measured observables are related linearly via Equations \ref{eq:x} and \ref{eq:y}, equivalent volumes in parameter space can be transformed by

\begin{equation}
\sigma_{X} \sigma_{Y} = \frac{R_o^2}{\left| \begin{array}{cc} B & G \\  A & F \\ \end{array} \right|} \sigma_x \sigma_y = f(\Omega, \omega, i) R_o^2 \sigma_x \sigma_y ,
\end{equation}
which reduces to $\sigma_{X}\sigma_{Y} \propto R_o^2 \sigma_x \sigma_y$ because the Jacobian is simply some function $f$ of the angular parameters. Analogously, $\sigma_{V_X}\sigma_{V_Y} \propto \sigma_{v_z}^2$, where the factor of $R_o^2$ does not appear because the observable $v_z$ is measured in units of physical distance per unit time, and thus there is no conversion from angular to physical distance. Plugging these relations into Equation \ref{eq:obs_prior}, we obtain a form of the two observable-based priors used in this paper:
\begin{equation}\label{eq:obsprior}
 \mathcal{P}(P, e) \propto 
 \left\{
 \begin{array}{ll}
  \sum_i \frac{1}{R_o^2\sigma_x(t_i)\sigma_y(t_i)} \left|J_{\rm{astro}}(t_i)\right| & \rm{(a)} \\
  \sum_i \frac{1}{R_o^2\sigma_x(t_i)\sigma_y(t_i)} \left|J_{\rm{astro}}(t_i)\right| + \sum_j\frac{1}{\sigma_{v_z}^2(t_j)} \left|J_{\rm{RV}}(t_j)\right|. & \rm{(b)}
 \end{array}
 \right.
\end{equation}

\section{Methodology and Observational Data}\label{sec:obsData}
We implement the new observable-based prior in {\em Efit5} \citep{Meyer2012}, an orbit fitting code developed by the {\em Galactic Center Orbits Initiative}. {\em Efit5} performs a Bayesian analysis using MULTINEST \citep{Feroz2008, Feroz2009}, a multi-modal nested sampling algorithm.

\begin{figure*}
\begin{center}
\includegraphics[width=\hsize]{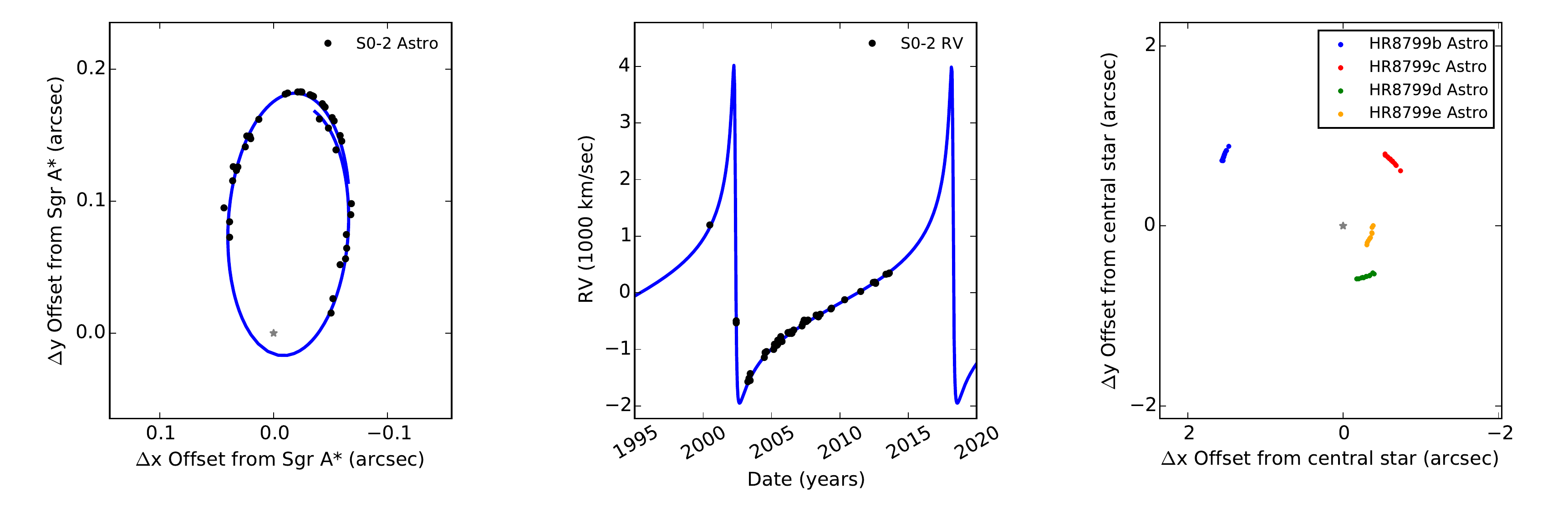}
\caption{\label{fig:obs_data} {\bf Left:} S0-2 Astrometric data published by \citet{Boehle2016}, with the best-fit orbit shown in blue. The gray star indicates the position of the central SMBH. The time baseline of observations extends from 1995 -- 2013, so although 2002 data (taken during the previous closest approach) are not plotted to avoid possible effects of confusion between S0-2 and the SMBH, there is full orbital phase coverage. {\bf Middle:} S0-2 RV data from the same source, with the best-fit RV curve shown in blue. {\bf Right:} HR 8799 astrometric data published by \citet{Konopacky2016}. The best-fit orbits are not plotted here since, with such low phase coverage, parameter posteriors are multi-modal. The gray star indicates the position of the central star.}
\end{center}
\end{figure*}

The following analysis uses astrometric and RV data published by \citet{Boehle2016} for the short-period star S0-2, and astrometric data published by \citet{Konopacky2016} for the HR 8799 directly-imaged exoplanet system. We use solely Keck HR 8799 data rather than data from multiple cameras and reduction pipelines to isolate the effects of phase coverage by eliminating the additional systematics originating from the use of multiple data sets \citep{Konopacky2016}. The model for the HR 8799 analysis assumes that the origin of the system is fixed and that the mass of and the line-of-sight distance to the host star are known. The mass of the star was set to 1.516 $M_{\odot}$ \citep{Baines2012} with an uncertainty of $\pm$ 0.15 $M_{\odot}$ \citep{Konopacky2016}, and the distance to the star was set to 39.4 $\pm$ 1.1 pc  \citep{VanLeeuwen2007}. A summary of the S0-2 and HR 8799 observations is provided in Table \ref{tab:planet_params}, and the data are plotted in Figure \ref{fig:obs_data}.

We test the performance of the observable-based prior compared to that of commonly-assumed uniform priors using simulations based on these data in Sect. \ref{sec:testing}, and apply the prior directly to these S0-2 and HR 8799 data in Sect. \ref{sec:application} to evaluate the scientific impact. 
\begin{table*}[h!]
\caption{\label{tab:planet_params}Summary of Observational Data}
\begin{center}
\begin{tabular}{@{}lccccc}
\noalign{\hrule height 1pt}
Object & $\delta\phi_{\mathrm{astro}}$ $^\mathrm{a}$ & $\delta\phi_{\mathrm{RV}}$ $^\mathrm{b}$ & Number of 		 &  Number of   & Data Source  \T\\
           & 							&							& Astrometry Points  & RV Points 	&    \B\\
\hline
S0-2         & $\sim$102\%  			&  $\sim$96\%& 38  &47 & \citet{Boehle2016} \T\\
HR 8799b & $\sim$3	\%				   & -- 		& 13 & -- & \citet{Konopacky2016}  \\
HR 8799c & $\sim$5\%				    & -- 		& 13 & -- &  \citet{Konopacky2016} \\
HR 8799d & $\sim$7\%				   & -- 		& 12 & -- &  \citet{Konopacky2016} \\
HR 8799e & $\sim$11\%				  & -- 		& 9  & -- &  \citet{Konopacky2016}  \B\\
\noalign{\hrule height 1pt}
\multicolumn{4}{@{}l}{\noindent \tiny $^\mathrm{a}$ Angular phase coverage: percentage of the 3-D orbit covered by astrometric observations, based on true anomaly.}\\
\multicolumn{4}{@{}l}{\noindent \tiny $^\mathrm{b}$ RV phase coverage: percentage of the 3-D orbit covered by RV observations, based on true anomaly.} \\
\end{tabular}
\\
\end{center}
\end{table*}

\section{Testing the Observable-based Prior}\label{sec:testing}
\subsection{Simulated Data Sets}\label{sec:simtests}
In this section, we describe the simulated data sets that are used to evaluate the effects of the observable-based prior on fitted model parameters. The simulated GC data are based on the short-period star S0-2 and exoplanet data are based on HR 8799d. The input parameter values from which mock data were generated are provided in Table \ref{tab:siminput}. One hundred simulations were run for each test case. Mock data were drawn from a normal Gaussian distribution with a mean equal to the value predicted by the model at each epoch and a standard deviation equal to the assumed error, as described below.

For the GC test cases, we simulated S0-2 data with varying angular phase coverage -- defined here as the percentage of the orbit that is swept by observations, with respect to the true anomaly. The simulations are based on a single star to keep variables such as angular orientation constant while only varying phase coverage, though future work will probe the effects of differences in intrinsic orbital parameters. To date, S0-2 is the brightest star in the central cluster with full orbital phase coverage. As such, S0-2's relatively well-defined orbit provides a realistic data set as a basis for comparison. A model orbit was created from a set of assumed parameter values published by \citet{Boehle2016}. All simulated data points assume an astrometric and RV error equal to S0-2's average observational error. 

Exoplanet mock data were generated in the same manner as were the S0-2 simulations, with sampling dates and average errors based on HR 8799d astrometry published by \citet{Konopacky2016}. Simulations were only run for one of the HR 8799 planets as an example because the statistical measures used to evaluate prior performance with these simulations are not as robust to multi-modality as that which is used to evaluate prior performance with the true data for these astrometry-only cases (see Sects. \ref{sec:sim_stats} and \ref{sec:EntropyStats}). 

Table \ref{tab:params} summarizes the simulated data sets. Test Case 1, which serves as a full-phase-coverage example, samples S0-2 mock data on the same dates for which astrometric and RV data were reported by \citet{Boehle2016}. For all subsequent GC test cases, evenly-spaced observations of S0-2 were simulated every 6 degrees on the plane of the sky ($\pm$ 0.5 degrees to simulate multiple observations per year). In Test Case 2, data points are centered in time on the periapse. The simulated observations in Test Case 2 cover $\sim$86\% of the orbit based on the true anomaly, but only $\sim$50\% of the orbit on the plane of the sky. All variations of Test Case 3 are centered on the apoapse with varying degrees of phase coverage but with the same observing cadence. This is achieved by systematically removing data points closest to the periapse for each subsequent simulation such that the total orbital phase coverage (and thus the number of data points) decreases without affecting the cadence of remaining measurements. Test Case 4 samples HR 8799d mock data on the same dates for which astrometric data were reported by \citet{Konopacky2016}.

\begin{table*}[h!]
\caption{\label{tab:siminput} Simulation input parameter values}
\begin{center}
\begin{tabular}{@{}lcc}
\noalign{\hrule height 1pt}
  & $^\mathrm{b}$S0-2 Simulations   & $^\mathrm{c}$HR 8799d Simulations               \\
\hline
Global Parameters$^\mathrm{a}$ && \\
\hline
$M_{\mathrm{cent}}$ (Solar Masses) 	  & 4.02 x $10^6$     &  1.516		\\	   
$R_o$ (pc)                                                    & 7860   		& 	39.4	 	\\	    
$x_o$ (mas) 						  & 2.74			& 	0	\\		 
$y_o$ (mas) 						  & -5.06			& 	0	\\		 
$v_{x,o}$ (mas yr$^{-1}$) 				  & -0.04			& 	0	\\		  
$v_{y,o}$ (mas yr$^{-1}$) 				  & 0.51			& 	0	\\		 
$v_{z,o}$  (km/s) 					  & -15.84			& 	0	\\				 
\hline		
Orbital Parameters$^\mathrm{a}$ && \\
\hline
 $\Omega$ (deg) 					& 228.0		& 59				 \\
 $\omega$  (deg) 					& 66.8		& 92.4		 \\
 i  (deg) 							& 134.2		& 29			\\
 P  (yr) 							& 15.92		& 112.7 		 \\
 $T_o$  (yr) 						& 2002.347	& 1995.4 	 \\  
 e 								& 0.892	  	& 0.02 	\\
\hline 
\noalign{\hrule height 1pt}
\multicolumn{3}{@{}l}{\noindent \tiny $^\mathrm{a}$ See Section \ref{sec:prior_orbit} for description of parameters.} \\
\multicolumn{3}{@{}l}{\noindent \tiny $^\mathrm{b}$ Input values for S0-2 simulations, derived from a simultaneous orbital fit of S0-2 and S0-38 with a jack knife bias term added for the reference frame  \citet{Boehle2016}.}\\
\multicolumn{3}{@{}l}{\noindent \tiny $^\mathrm{c}$ Input values for HR 8799d simulations, based on values used for simulations from \citet{Konopacky2016}. }\\
\end{tabular}
\end{center}
\end{table*}

\begin{table*}[h!]
\caption{\label{tab:params}Summary of simulated test cases}
\begin{center}
\begin{tabular}{@{}lccccc}
\noalign{\hrule height 1pt}
Test Case$^\mathrm{a}$  & $\delta\phi_{\mathrm{astro}}$ $^\mathrm{b}$ & $\delta\phi_{\mathrm{RV}}$ $^\mathrm{c}$ & Number of  &  Number of &  Description\T\\
   &  &  & Astrometry Points  &  RV Points & \B\\
\hline
1        &   101.9\% &   95.8\%  & 38 &  47 &  S0-2 True Sampling   \T\\
2        &    86.0\%  &   86.0\%  & 18 &  18 &  S0-2 Periapse Centered  \\
3.1     & 	71.9\% &    71.9\% &  56 & 56  &  S0-2 Apoapse Centered	  \\
3.2     &     66.5\%  &   66.5\%  &  55 & 55 &       \\
3.3     &     41.4\%  &   41.4\%  &  53 & 53 &       \\                     
3.4     &     31.7\%  &   31.7\%  &  52 & 52 &       \\               
3.5     &     27.2\%  &   27.2\%  &  51 & 51 &       \\                       
3.6     &     25.9\%  &   25.9\%  &  51 & 51 &       \\         
3.7    &      21.4\%  &   21.4\%  &  50 & 50 &       \\                    
3.8    &      18.3\%  &   18.3\%  &  48 & 48 &       \\      
3.9    &      16.2\%  &   16.2\%  &  45 & 45 &       \\             
3.10  &      15.3\%  &   15.3\%  &  46 & 46 &       \\                        
3.11  &      13.9\%  &   13.9\%  &  44 & 44 &       \\                       
3.12  &      13.7\%  &   13.7\%  &  42 & 42 &       \\                      
3.13  &      11.1\%  &   11.1\%  &  40 & 40 &       \\
4       &      7.2\%  &   --  &  12 & -- &  HR 8799d True Sampling    \B\\
\noalign{\hrule height 1pt}
\multicolumn{4}{@{}l}{\noindent \tiny $^\mathrm{a}$ See Sect. \ref{sec:simtests} for a description of the simulated data sets.}\\
\multicolumn{4}{@{}l}{\noindent \tiny $^\mathrm{b}$ Angular phase coverage: percentage of the 3-D orbit covered by astrometric observations, based on true anomaly.}\\
\multicolumn{4}{@{}l}{\noindent \tiny $^\mathrm{c}$ RV phase coverage: percentage of the 3-D orbit covered by RV observations, based on true anomaly.}\\  
\end{tabular}
\end{center}
\end{table*}

\subsection{Simulation Results and Statistical Measures}\label{sec:sim_stats}
For the simulations described in Sect. \ref{sec:simtests}, two statistical measures are used to evaluate the general performance of our observable-based prior compared to that of uniform priors. We define a new statistic, the Bias Factor, as well as use a classical statistic, the statistical efficiency, to quantify how well regions near the true value are sampled. Consistently sampling regions of parameter space that are favored by the data is one property that must be satisfied for posteriors to be truly objective \citep{Bernardo2005}. Although observable-based priors are weakly-informative rather than truly objective, we show that the less-subjective nature of these priors (compared to that of commonly-assumed uniform priors) indeed promotes consistent sampling by showing that it improves the Bias Factor and statistical efficiency. 

\subsubsection{Bias Factor} \label{sec:BiasStats}
Given a hypothetical large set of N orbital fits resulting from N mock data sets, we define the Bias Factor $\beta_{\mathrm{F}}$ as
\begin{equation}\label{eq:BiasFactor}
\beta_{\mathrm{F}} = \underset{{\mathrm{i}}\in {\mathrm{N}}} {\mathrm{median}} \left\{\beta_{\mathrm{i}}\right\},
\end{equation}
where $\beta_{\mathrm{i}}$ is defined to be the Specific Bias in parameter $\theta$ for the $i-$th orbital fit. Here, $\theta$ denotes any parameter in the set of model parameters, $\mathscr{M}$. The Specific Bias $\beta_{\mathrm{i}}$ is given by the difference between the median parameter value ($\hat{\theta}_{\mathrm{i}}$) and the assumed true value ($\theta_{\mathrm{true}}$), normalized by half the 68\% central credible interval for that parameter ($\sigma_{\mathrm{i}}$, see Sect. \ref{sec:EfficiencyStats}):
\begin{equation}\label{eq:Bias}
\beta_{\mathrm{i}} = \frac{ \hat{\theta_{\mathrm{i}}} - \theta_{\mathrm{true}} }{\sigma_{\mathrm{i}}}.
\end{equation}

The Specific Bias effectively measures the deviation from the assumed true parameter value, in units of $\sigma$. Ideally, for each orbital fit, $\hat{\theta}$ has a fifty percent probability of being greater than $\theta_{\mathrm{true}}$, and a fifty percent probability of being less than $\theta_{\mathrm{true}}$. If no bias were present, the probability distribution of $\beta_i$ values would be normally distributed about zero. For a large number of data sets, the Bias Factor should thus, in theory, be consistent with zero. A Bias Factor greater (less) than zero indicates that the parameter is statistically biased high (low).

\begin{figure*}
\begin{center}
\includegraphics[width=\hsize]{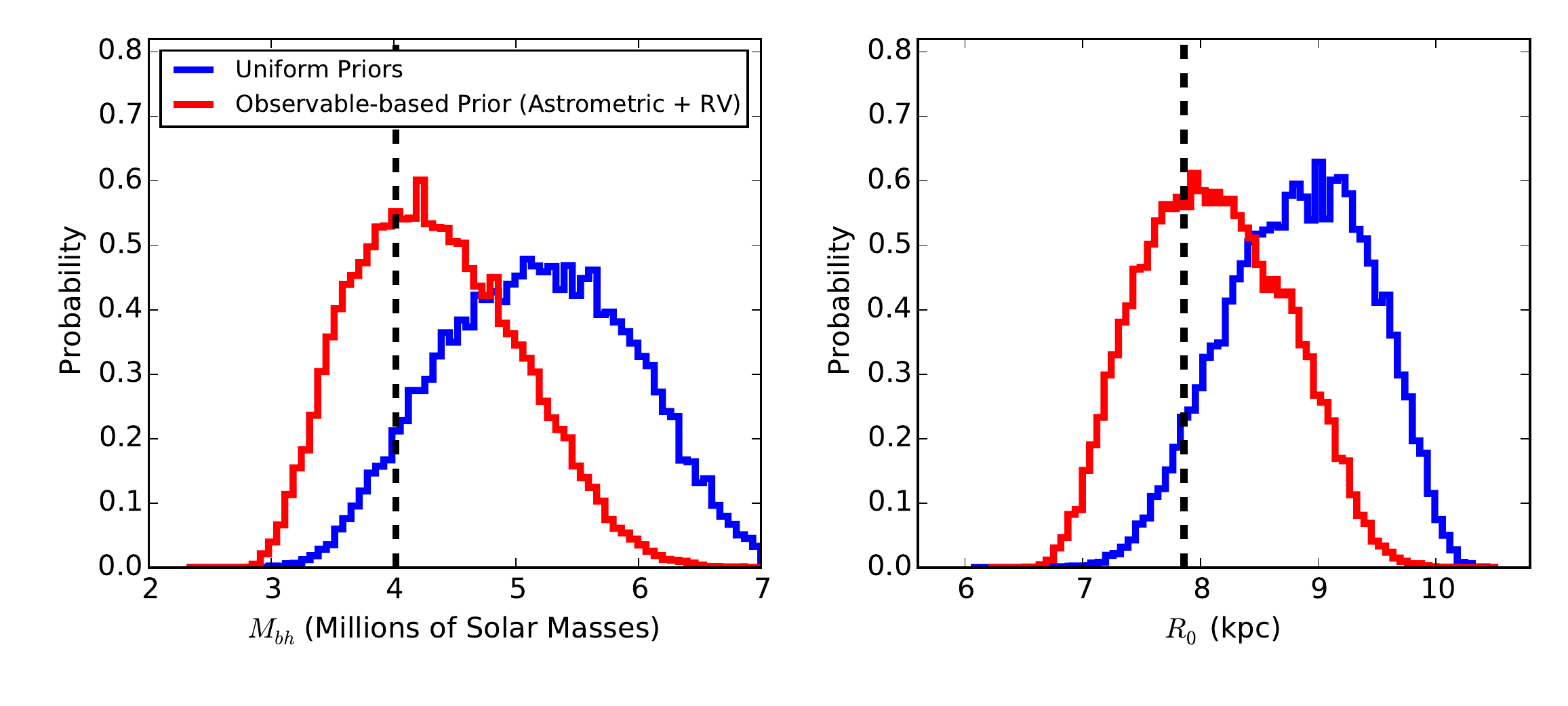}
\caption{\label{fig:MRposterior} Marginalized 1-D Posteriors for the black hole mass (left) and the line-of-sight distance to the GC (right) from a single orbital fit for Test Case 3.9, $\sim$16\% phase coverage centered on the apoapse, assuming uniform priors (blue) and a new prior based on uniformity in the astrometric and RV observables (red). All curves are normalized such that the area under the curve is equal to one. For reference, the assumed true value from which the mock data are generated is indicated by the dashed vertical line. The observable-based prior, as compared to standard uniform priors, allows the posteriors to shift closer to the assumed true value, indicating a less biased result (smaller specific bias value).}
\end{center}
\end{figure*}

\begin{figure*}
\begin{center}
\includegraphics[width=\hsize]{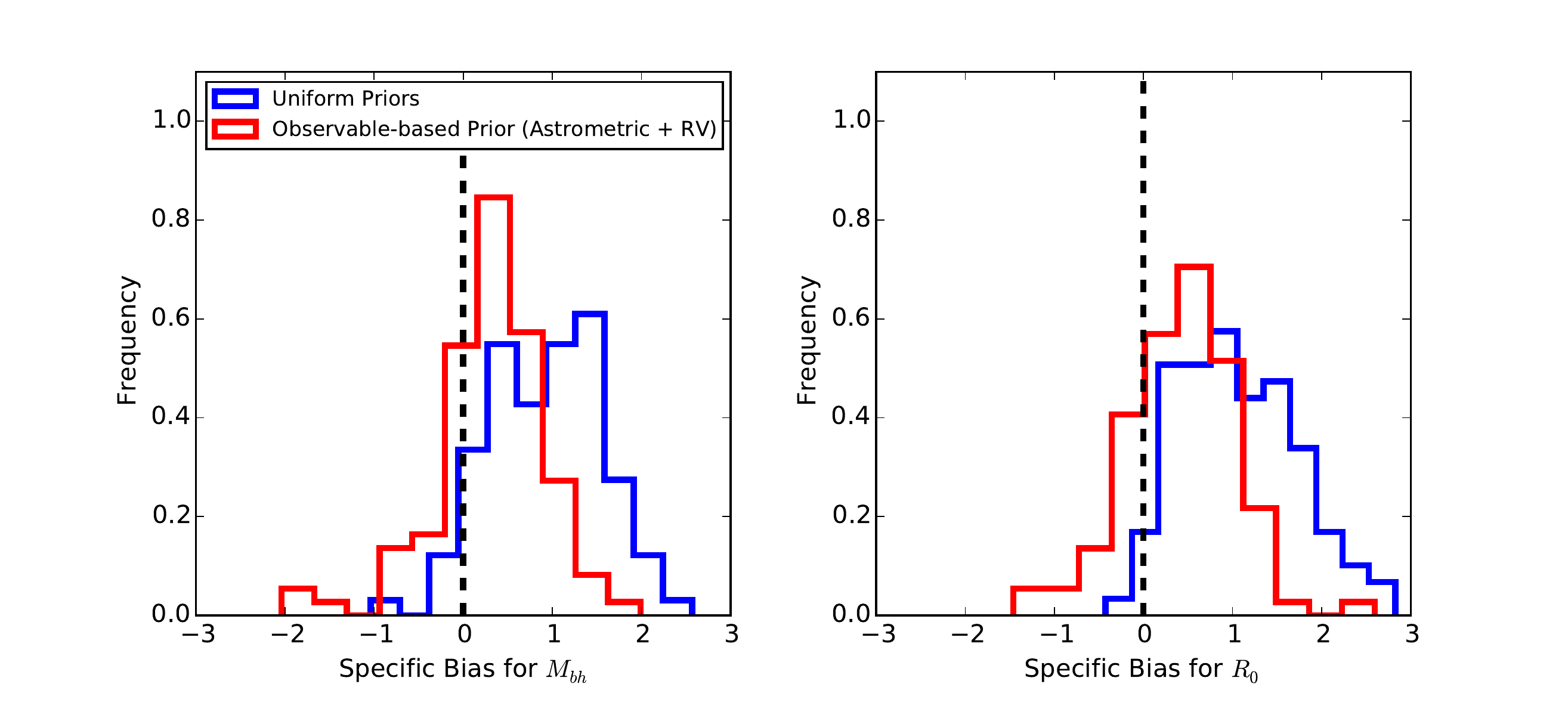}  
\caption{\label{fig:SpecificBias} Probability distribution of Specific Bias values (Eq. \ref{eq:Bias}) for the set of N = 100 orbit fits in Test Case 3.9, $\sim$16\% phase coverage centered on the apoapse. The distributions are plotted for the black hole mass (left) and the line-of-sight distance to the GC (right) assuming uniform priors (blue) and a new prior based on uniformity in the astrometric and RV observables (red). All curves are normalized such that the area under the curve is equal to one. The dashed line at zero indicates an unbiased result. The distribution shifts closer to zero with the observable-based prior as compared to the uniform prior, indicating that the reduction in bias seen in Figure \ref{fig:MRposterior} is statistical. The Bias Factor is then defined as the median of each of the above distributions.}
\label{fig:specific_pdf}
\end{center}
\end{figure*}

We first illustrate the Specific Bias by looking at the effects of the prior on a single simulation. Figure \ref{fig:MRposterior} shows marginalized 1-D $M_{\mathrm{bh}}$ and $R_o$ posteriors for a randomly chosen fit in Test Case 3.9 (Table \ref{tab:params}), which serves as a low-phase-coverage example.  The resulting posteriors shift closer to the assumed true value when using an observable-based prior rather than uniform priors, indicating a Specific Bias value closer to zero (less biased). Figure \ref{fig:SpecificBias} shows the set of $\beta_i$ values for all 100 orbit fits for the same test case to confirm that the improvement detected in Figure \ref{fig:MRposterior} is significant. Here, the distribution of $\beta_i$ values shifts closer to zero with the observable-based prior. 

\begin{figure*}
\begin{center}
\includegraphics[width=\hsize]{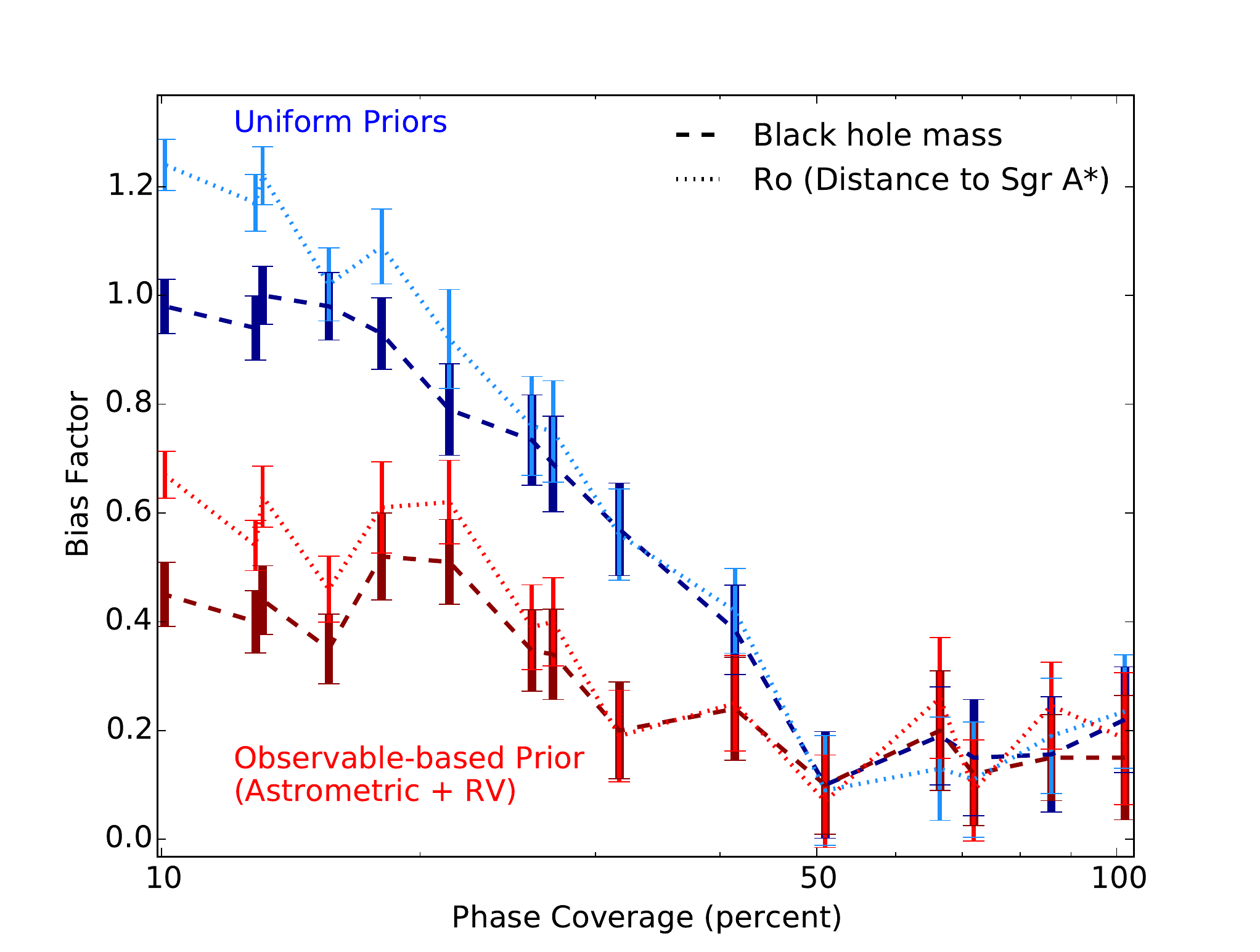}
\caption{\label{fig:phaseVSbias} Bias Factor (Section \ref{sec:BiasStats}) as a function of phase coverage for the black hole mass and the line-of-sight distance to the GC, assuming uniform priors (blue) and a new prior based on uniformity in the astrometric and RV observables (red). Error bars indicate 1-$\sigma$ standard errors on the Bias Factor. The onset of bias occurs at less than $\sim$40\% phase coverage, and is less dramatic with the observable-based prior than with uniform priors. At low phase coverage, there is a factor of two improvement in the Bias Factor with the observable-based prior over standard uniform priors.} 
\end{center}
\end{figure*}

We quantify this shift towards a less-biased result by evaluating the Bias Factor, $\beta_{\mathrm{F}}$. Figure \ref{fig:phaseVSbias} shows the Bias Factor for $M_{\mathrm{bh}}$ and $R_o$ for all GC test cases as a function of phase coverage. As phase coverage decreases, the prior has a more profound impact on parameter estimates. When more than half of the orbit (based on true anomaly) has been observed, the Bias Factor is low and remains roughly consistent between the uniform prior and the observable-based prior. Below $\sim$40\% phase coverage, the Bias Factor increases more rapidly with uniform priors than it does with the observable-based prior. This cutoff is consistent with the onset of bias seen by \citet{Lucy2014}. With the observable-based prior, the Bias Factor improves by a factor of two at low phase coverage. For example, the Bias Factor for $R_o$ rises to greater than 1.2 with uniform priors, but remains $\sim$0.6 with the observable-based prior. Similarly, the Bias Factor for $M_{\mathrm{bh}}$ rises to greater than 1.0 with uniform priors, but remains less than 0.5 with the observable-based prior. If the Bias Factor for a given parameter is 1, then the average output value of that parameter is 1-$\sigma$ greater than the true value. A Bias Factor of 1 would indicate a consistent shift from the ``true'' value that can have profound effects on the accuracy of resulting inferences. 

\begin{figure*}
\begin{center}
\includegraphics[width=\hsize]{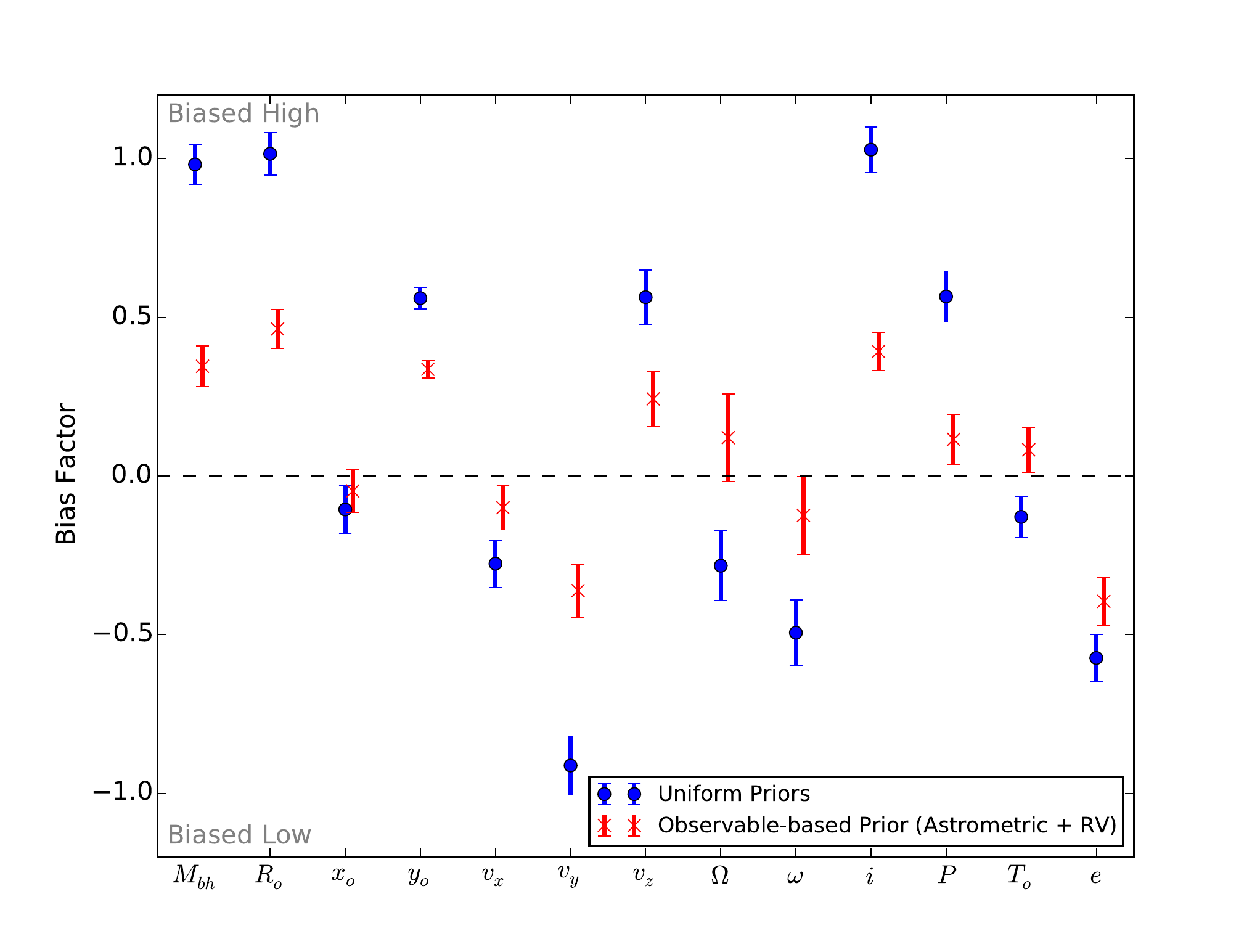}
\caption{\label{fig:OrbitBias_lowcov}  Bias Factor (Section \ref{sec:BiasStats}) for all parameters for Test Case 3.9, which serves as an example of low angular phase coverge ($\sim$16\%), assuming uniform priors (blue) and a new prior based on uniformity in the astrometric and RV observables (red). Error bars indicate 1-$\sigma$ standard errors on the Bias Factor. The Bias Factor is reduced for all parameters by assuming an observable-based prior rather than standard uniform priors.}  
\end{center}
\end{figure*}

Although we only highlight the Bias Factor for $M_{\mathrm{bh}}$ and $R_o$ in Figure \ref{fig:phaseVSbias} for clarity, the Bias Factor is improved with the observable-based prior for all global and orbital parameters in cases of low phase coverage. Figure \ref{fig:OrbitBias_lowcov} shows this improvement in the Bias Factor for all parameters for Test Case 3.9, which again serves as an example of a low-phase-coverage test. In this case, by assuming an observable-based prior rather than uniform priors, the Bias Factor decreases from 0.98 $\pm$ 0.06 to 0.35 $\pm$ 0.06 for $M_{\mathrm{bh}}$ (a 65\% improvement), and from 1.02 $\pm$ 0.07 to 0.46 $\pm$ 0.06 for $R_o$ (a 54\% improvement). There are similar improvements in all parameters in this case -- a 55\% improvement in $x_o$, 40\% in $y_o$, 64\% in $v_{x,o}$, 61\% in $v_{y,o}$, 57\% in $v_{z,o}$, 57\% in $\Omega$, 75\% in $\omega$, 62\% in $i$, 80\% in $P$, 36\% in $T_o$, and 31\% in $e$. 

Because credible intervals are difficult to define for multi-modal distributions, data must have some minimum constraining power for the Bias Factor to be a robust performance metric. With greater than $\sim$10\% phase coverage and with both astrometric and RV data, resultant posteriors are mono-modal and we can define a central credible interval for the GC cases presented. We do not present the Bias Factor for GC test cases with less than $\sim$10\% phase coverage because the errors become so large that nearly the entire parameter range is covered and the utility of the Bias Factor as a performance metric breaks down. Similarly, in astrometry-only cases such as HR 8799, posteriors are often multi-modal without the additional constraints provided by RV data, causing central credible intervals not to be well-defined. We therefore do not run simulations for all HR 8799 planets since the applicable analyses depend on central credible interval construction. However, it is interesting to evaluate the statistical properties of eccentricity estimates nonetheless because low phase coverage is known to bias eccentricity estimates towards artificially high values \citep[e.g.][]{Konopacky2016}.  As an example, we run simulations for HR 8799d, and evaluate the eccentricity bias using the 68\% error on the mode rather than central credible intervals. Eccentricity estimates indeed are biased high when uniform priors are assumed, but are consistent with the input value with our new observable-based prior. For HR 8799d, the Bias Factor on $e$ is reduced from $3.17 \pm 0.21$ with uniform priors to an unbiased value of $0.07 \pm 0.26$ with the observable-based prior. This result indicates that using the observable-based prior with low-phase-coverage data can help mitigate the known risk of biasing eccentricity estimates towards artificially high values. We discuss the reason for this improvement in Sect. \ref{sec:HRresults}.

\subsubsection{Statistical Efficiency}\label{sec:EfficiencyStats}
We use the statistical efficiency to investigate how the observable-based prior affects the reliability of calculated confidence intervals. Bayesian confidence intervals (credible intervals) and classical confidence intervals are fundamentally different in both construction and interpretation \citep{Host2007}. Unlike credible intervals, which are defined simply as regions containing a prescribed posterior probability, classical confidence intervals have a more intricate definition.  In the classical definition, for a sufficiently large number of experiments, the confidence interval inferred from each experiment will contain, or cover, the universally ``true'' value a prescribed fraction of the time (confidence level $\times$ 100\%) \citep{Neyman1937}. By this definition, a 68\% confidence interval requires that 68 out of 100 possible observed (or randomly drawn) datasets produce a confidence interval that covers the true value. Confidence intervals that cover the true value more than the prescribed frequency are said to over-cover, whereas confidence intervals that cover the true value less than the prescribed frequency are said to under-cover. Unfortunately, most algorithms used to calculate confidence intervals do not guarantee exact coverage, and thus can either under- or over-cover. Under- or over-covering is common in prior-dominated regimes where data are not rigorously constraining. In such cases, prior information can have a profound impact on the resulting confidence intervals \citep{Lucy2014}. 

Statistical efficiency is a powerful performance diagnostic that is used to investigate the accuracy of calculated confidence or credible intervals. Statistical efficiency is defined as the ratio of effective coverage (the experimentally determined percentage of datasets in which the inferred confidence or credible interval covers the true value) to stated coverage (68\% for a 1-$\sigma$ confidence interval). While based on the classical definition of confidence intervals, effective coverage can be used to evaluate the effectiveness of both credible and confidence intervals \citep{Cameron2011}. An efficiency of 1 indicates exact coverage, whereas an efficiency less (greater) than 1 indicates that the credible interval under-(over-)covers.

\begin{figure*}
\begin{center}
\includegraphics[width=\hsize]{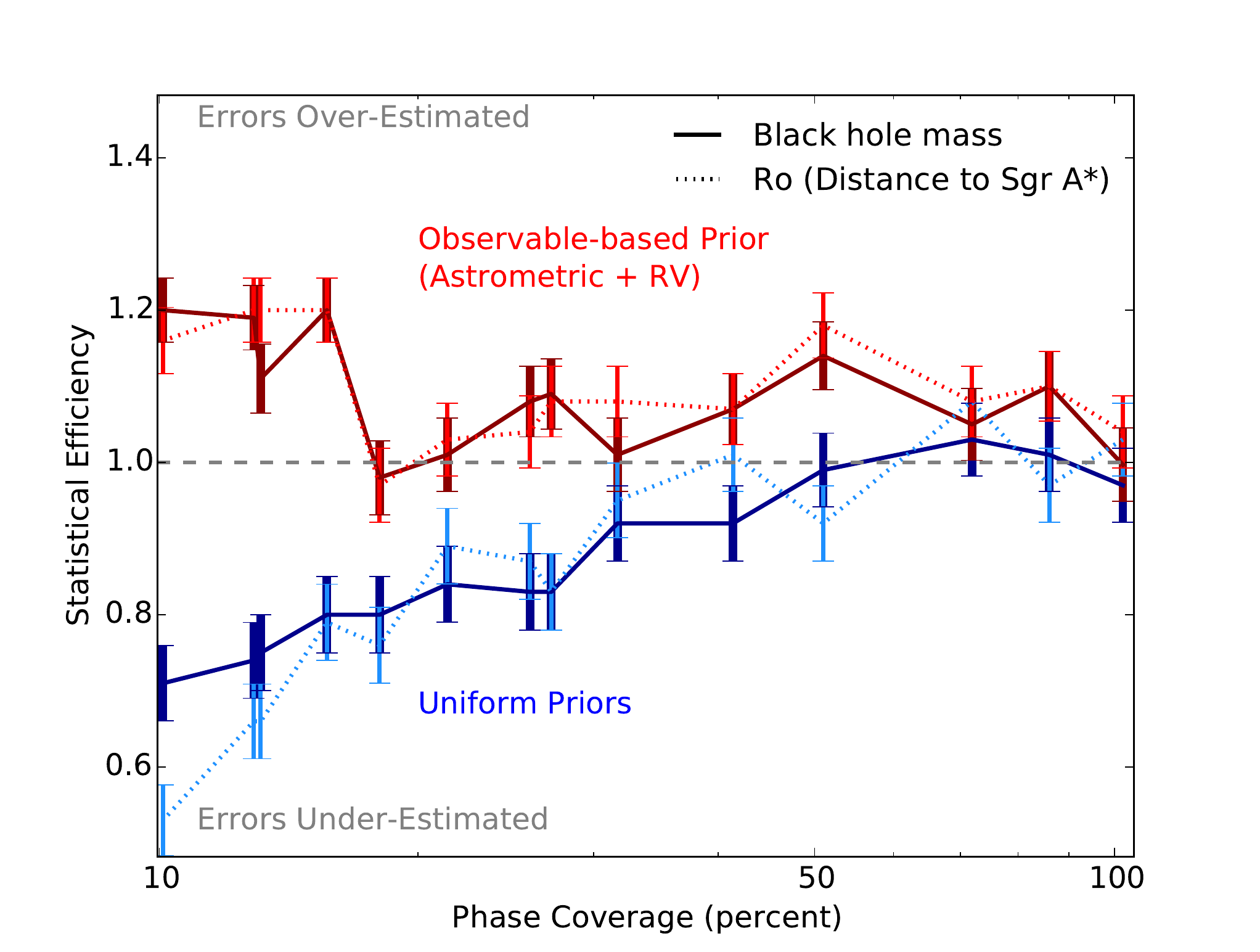}
\caption{Statistical efficiency (Section \ref{sec:EfficiencyStats}) as a function of phase coverage for the black hole mass and the line-of-sight distance to the GC, assuming a uniform prior (blue) and a new prior based on uniformity in the astrometric and RV observables (red). Errors on the statistical efficiency are derived from a binomial distribution (with N = 100 trials). The dashed horizontal line at one indicates exact coverage. Below $\sim$40\% phase coverage, the statistical efficiency decreases to less than one when assuming a uniform prior, indicating that credible intervals under-cover (errors are under-estimated). Credible intervals remain well-defined when assuming an observable-based prior as the statistical efficiency remains at or greater than one. In this case, the observable-based prior eliminates the problem of under-estimated errors.}
\label{fig:phaseVSefficiency}
\end{center}
\end{figure*}

Figure \ref{fig:phaseVSefficiency} presents the statistical efficiency as a function of phase coverage for all simulated GC test cases, again focusing on the black hole mass ($M_{\mathrm{bh}}$) and the line-of-sight distance to the GC ($R_o$) for clarity. With phase coverage greater than $\sim$40\%, credible intervals are well-defined as the statistical efficiency is consistent with one for both $M_{\mathrm{bh}}$ and $R_o$, indicating nearly exact coverage. Below $\sim$40\% phase coverage, the statistical efficiency decreases to less than one with uniform priors, indicating that the derived credible intervals under-cover and errors are under-estimated. This drop in statistical efficiency is consistent with the onset of bias around 40\% phase coverage (Sect. \ref{sec:BiasStats}). The risk of under-estimating errors below $\sim$40\% phase coverage is mitigated by assuming an observable-based prior. The statistical efficiency remains greater than or equal to one for all S0-2 test cases with the observable-based prior. With statistical efficiencies greater than one, the credible intervals over-cover and error estimates are over-conservative. While exact coverage is ultimately desired, it is better to be over-conservative than to under-estimate the errors. 

In light of the predicted and detected eccentricity biases for HR 8799d, we also investigate the HR 8799d eccentricity efficiency. As with the Bias Factor, statistical efficiency depends on credible interval construction, and thus must be evaluated cautiously for cases such as HR 8799 where posteriors are multi-modal and central credible intervals are not always well-defined. As such, we again use the 68\% error on the mode of the posterior distribution rather than using central credible intervals. For HR 8799d, statistical efficiency for eccentricity increases from 0.6 $\pm$ 0.05 (under-covering) to greater than one (sufficiently-defined credible intervals) with the observable-based prior, indicating that the errors on the mode are no longer under-estimated with the new prior. 

\section{Application of the Observable-based Prior}\label{sec:application}
We now apply the observable-based prior to real S0-2 and HR 8799 data, described in Sect \ref{sec:obsData}, to evaluate the scientific impact.
\subsection{S0-2 Results}\label{sec:S02results}
Table \ref{tab:priors} shows that with full orbital phase coverage, the orbital solution for S0-2 derived with the observable-based prior is consistent with that published by \citet{Boehle2016}. S0-2 astrometric and RV data have high constraining power and thus the orbital solution is not influenced heavily by prior effects. This result is supported by simulation results for the high-phase-coverage limit in Figures \ref{fig:phaseVSbias} and  \ref{fig:phaseVSefficiency}. There are many stars with low phase coverage in the GC whose orbital solution could be improved by the observable-based prior; as such, future work includes using the new prior to fit orbits to these stars to probe the effects of the prior on stars with different eccentricity distributions and angular orientations.

\begin{table*}[h!]
\caption{\label{tab:priors}Orbital parameter estimates for S0-2 derived with the observable-based prior compared to previously published values}
\begin{center}
\begin{tabular}{@{}lccc}
\noalign{\hrule height 1pt}
  & $^\mathrm{b}$Prior Ranges   & $^\mathrm{c}$Observable-based Prior & $^\mathrm{d}$\citet{Boehle2016}                  \\
\hline
Global Parameters$^\mathrm{a}$ &&& \\
\hline
$M_{\mathrm{bh}}$ ($10^6$ Solar Masses) & [2.3 , 7.0]            &4.01 $\pm$ 0.31		&4.03 $\pm$ 0.31		\\	   
$R_o$ (kpc)                                                  & [5.90 , 10.5]        &7.98 $\pm$ 0.36		&8.01 $\pm$ 0.36		\\	    
$x_o$ (mas) 						  & [-6, 8] 			&2.04 $\pm$ 0.56		&2.02 $\pm$ 0.56		\\		 
$y_o$ (mas) 						  & [-8 , 8] 			&-3.70 $\pm$ 1.34		&-3.64 $\pm$ 1.32		\\		 
$v_{x,o}$ (mas yr$^{-1}$) 				  & [-0.3 , 0.8] 		&-0.10 $\pm$ 0.03		&-0.10 $\pm$ 0.03		\\		  
$v_{y,o}$ (mas yr$^{-1}$) 				  & [-0.5 , 1.5] 		&0.71 $\pm$ 0.07		&0.72 $\pm$ 0.07		\\		 
$v_{z,o}$  (km/s) 					  & [-120. , 120.] 	&-20 $\pm$ 10			&-19 $\pm$ 10			\\				 
\hline
Orbital Parameters$^\mathrm{a}$ &&& \\
\hline
 $\Omega$ (deg) 					& [221 , 233] 		& 227.9 $\pm$ 0.8		& 227.9 $\pm$ 0.8			 \\
 $\omega$  (deg) 					& [60 , 71] 		& 66.6 $\pm$ 0.9		& 66.5 $\pm$ 0.9		 \\
 i  (deg) 							& [128 , 139]  		& 134.6 $\pm$ 0.9		& 134.7 $\pm$ 0.9		\\
 P  (yr) 							& [15.4 , 16.9] 		& 15.90 $\pm$ 0.04		& 15.90 $\pm$ 0.04		 \\
 $T_o$  (yr) 						& [2002.29 , 2002.38] & 2002.344 $\pm$ 0.008  & 2002.343 $\pm$ 0.008	 \\  
 e 								& [0.865 , 0.915] 	& 0.890 $\pm$ 0.005 	& 0.890 $\pm$ 0.005 	\\
\hline 
\noalign{\hrule height 1pt}
\multicolumn{4}{@{}l}{\noindent \tiny $^\mathrm{a}$ See Section \ref{sec:prior_orbit} for description of parameters.} \\
\multicolumn{4}{@{}l}{\noindent \tiny $^\mathrm{b}$ Prior ranges for the observable-based prior are still specified in model-parameter space.}\\
\multicolumn{4}{@{}l}{\noindent \tiny $^\mathrm{c}$ Mean posterior estimates for S0-2 derived using same data as in column 4, but with the observable-based prior.}\\
 \multicolumn{4}{@{}l}{\hspace{1mm} \tiny  All results with the new prior are consistent with previously published values, as expected with full phase coverage.}\\
\multicolumn{4}{@{}l}{\noindent \tiny $^\mathrm{d}$ Mean posterior estimates reported by \citet{Boehle2016} derived from an individual orbital fit of S0-2, without the jack knife bias term added for the reference frame.}\\
\multicolumn{4}{@{}l}{\hspace{1mm} \tiny Though \citet{Boehle2016} also report values derived from a simultaneous fit of S0-2 and S0-38, we compare the values for S0-2 alone here for simplicity.}\\
\end{tabular}
\end{center}
\end{table*}

\subsection{HR 8799 Results and Statistical Analysis}\label{sec:HRresults}
\begin{figure*}
\begin{center}
\includegraphics[width=\hsize]{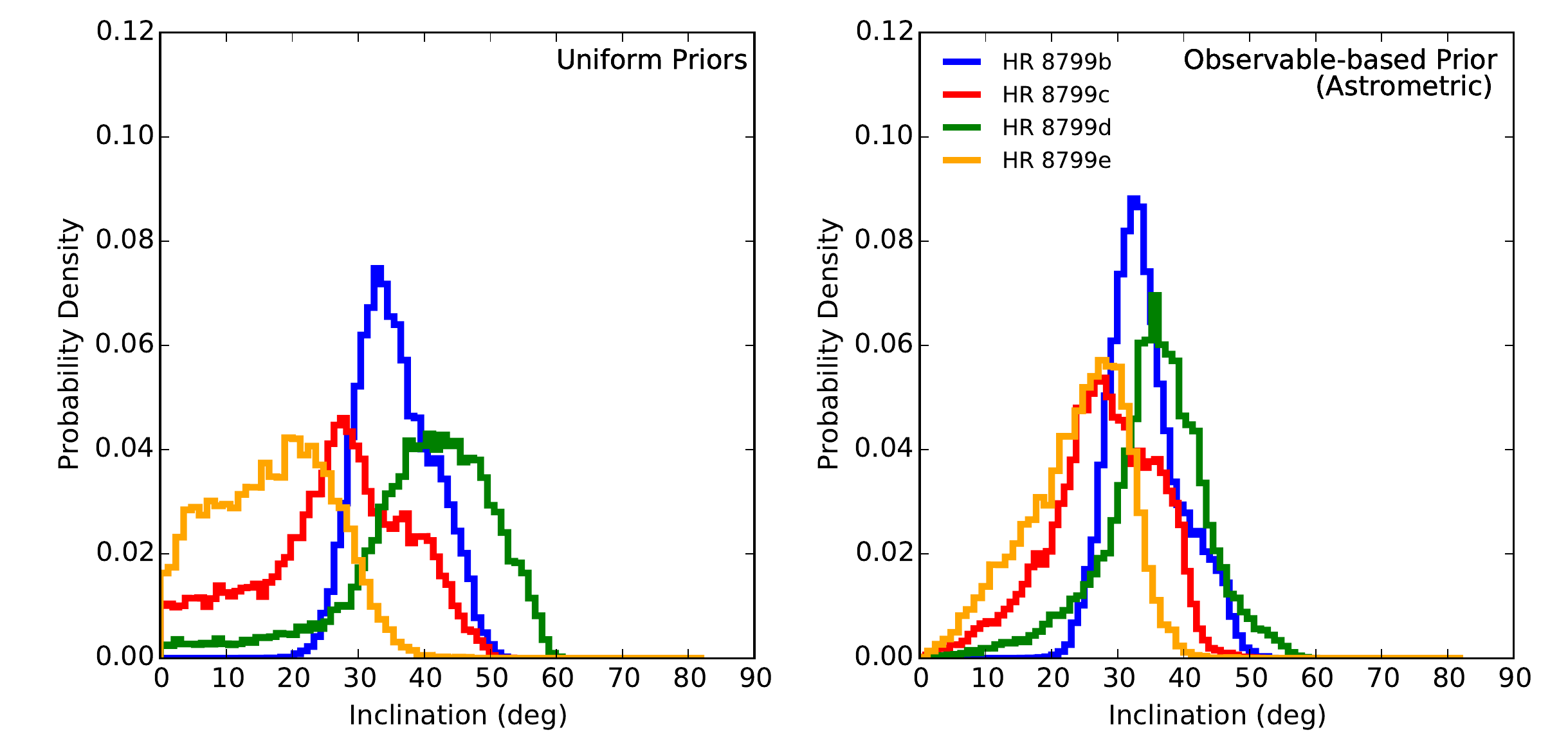}
\caption{\label{fig:inc_posteriors}  Inclination posteriors for HR 8799b, c, d, and e assuming uniform priors (left) and a new prior based on uniformity in the astrometric observables (right). The observable-based prior provides stronger evidence that the inclinations of the four planets are consistent around 30$^o$ to within 1-$\sigma$.}
\end{center}
\end{figure*}

\begin{figure*}
\begin{center}
\includegraphics[width=\hsize]{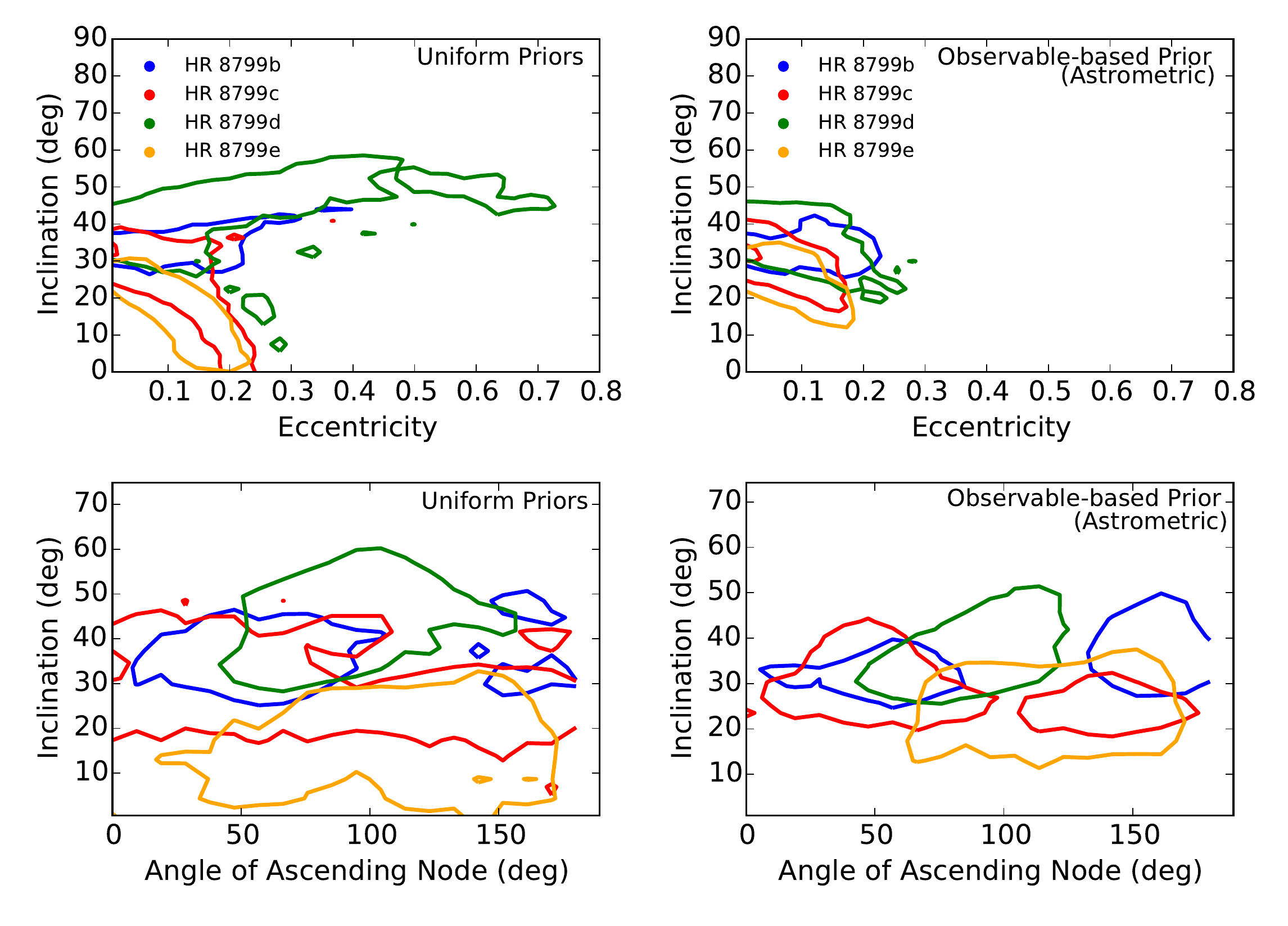}
\caption{\label{fig:eVSi_contour}  1-$\sigma$ contours for the joint probability distribution functions between eccentricity and inclination (top) and between angle of ascending node and inclination (bottom) for HR 8799b, c, d, and e assuming uniform priors (left) and a new prior based on uniformity in the astrometric observables (right).  With the observable-based prior, low eccentricity orbits are favored for all four planets with an inclination of $\sim$30$^\mathrm{o}$ to within 1-$\sigma$, placing stronger constraints on the orbital plane configuration of the HR 8799 system. The bottom panel shows that coplanar solutions are still allowed, though not necessitated, with the observable-based prior. Using the same data set, prior choice can influence our physical constraints on the system, highlighting the fact that prior effects must not be ignored.}
\end{center}
\end{figure*}

We investigate how the inferred orbital plane configurations of the HR 8799 planets change based on prior choice. Prior choice affects whether the planets are hypothesized to have consistent inclinations, and thus prior influences must be taken into consideration in this low-phase-coverage regime. The observable-based prior provides stronger evidence that the four planets have a consistent inclination of $\sim$30$^\mathrm{o}$ to within 1-$\sigma$ (see inclination posteriors in Figure \ref{fig:inc_posteriors}). Figure \ref{fig:eVSi_contour} shows 1-$\sigma$ contours for the joint probability distribution functions between eccentricity and inclination and between angle of ascending node ($\Omega$) and inclination for each of the four HR 8799 planets. While we cannot claim coplanarity between the host star and the planets because the $\Omega$ posteriors remain largely unconstrained, the possibility is allowed by this analysis (bottom panel of Figure \ref{fig:eVSi_contour}). Table \ref{tab:HRinc} lists the median values of the inclination posteriors derived with both flat priors and the observable-based prior for each of the four planets. 

\begin{figure*}
\begin{center}
\includegraphics[width=0.48\hsize]{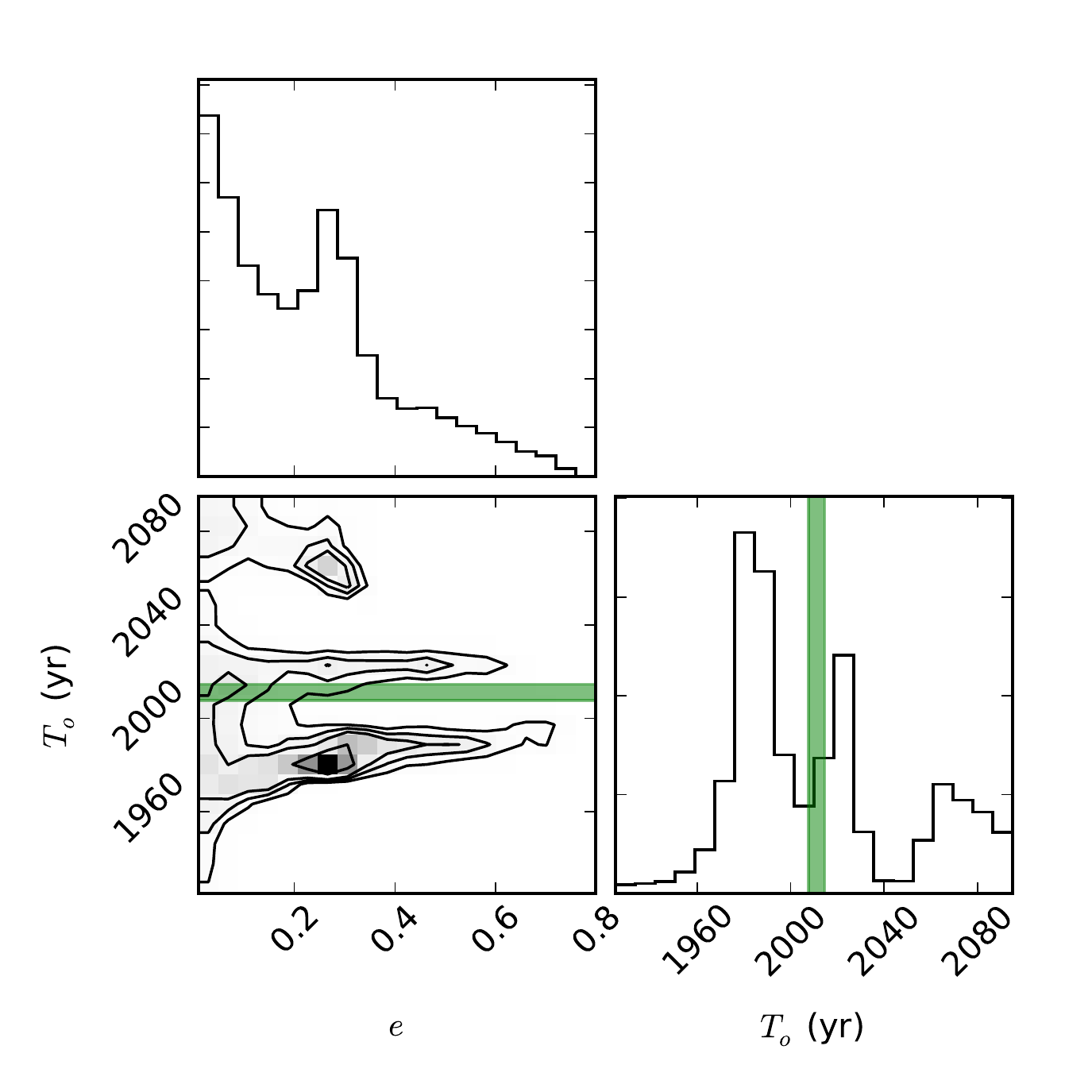}
\includegraphics[width=0.48\hsize]{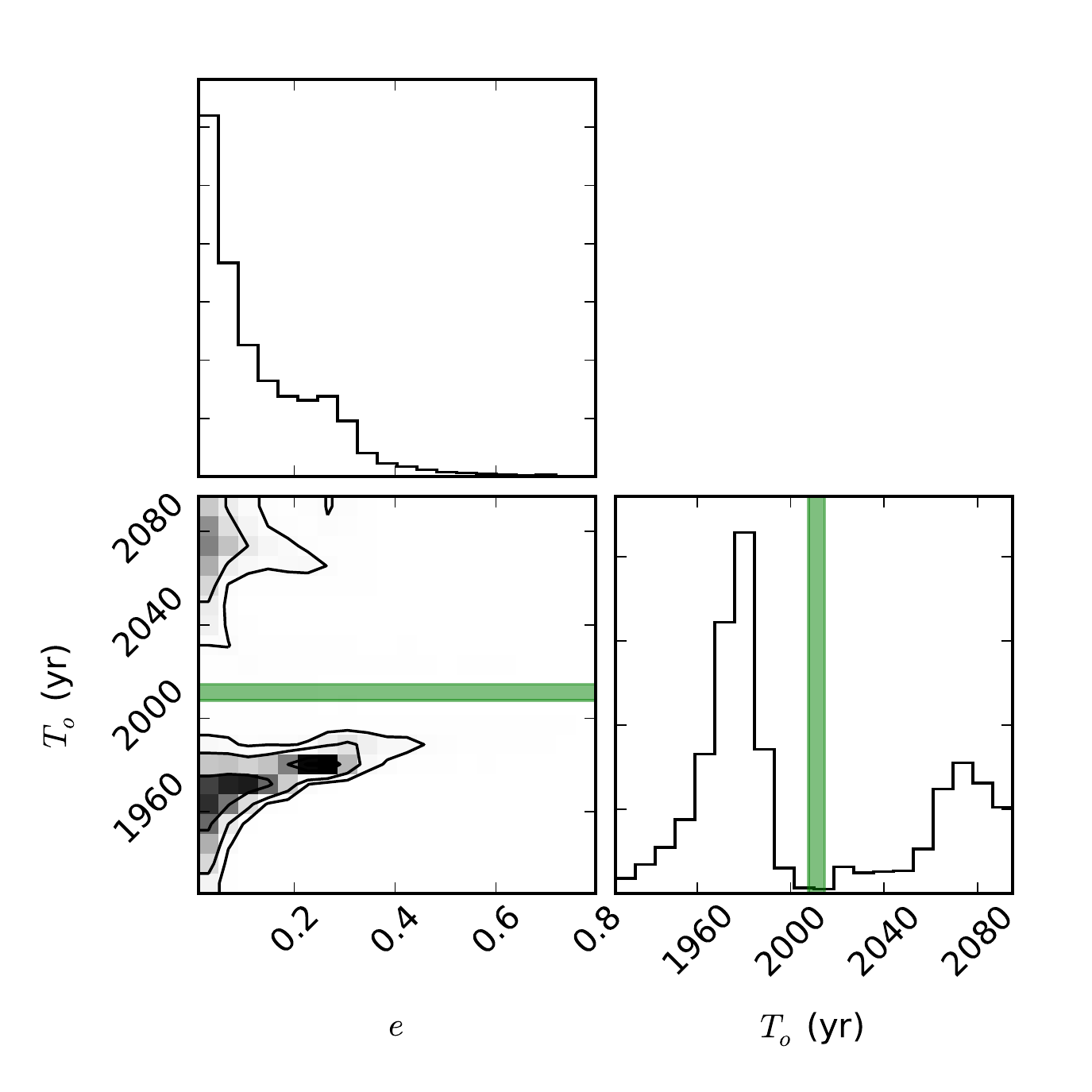}
\caption{\label{fig:Te_corner} Joint probability distribution functions between eccentricity and $T_o$ assuming uniform priors (left) and an observable-based prior (right) for HR 8799d. With uniform priors, an artificial $T_o$ mode emerges near the epochs during which observations were taken, indicated by the green shaded region. The left panel indicates that with uniform priors, this biased region of $T_o$ parameter space corresponds with an artificially high eccentricity mode. The observable-based prior mitigates this bias by suppressing the artificial $T_o$ mode and consequently the artificially high eccentricity mode. }
\end{center}
\end{figure*}

Further, the observable-based prior favors lower eccentricity estimates for all four planets than those inferred with uniform priors, providing more stringent eccentricity upper limits than those inferred with uniform priors and allowing for the possibility of nearly circular orbits (top panel of Figure \ref{fig:eVSi_contour}). This result is consistent with the HR 8799d simulation reported in Sect. \ref{sec:BiasStats}, which shows that HR 8799d eccentricity estimates are biased high with uniform priors, but are unbiased with the observable-based prior. Previous works have noted that such artificially high eccentricities can result from $T_o$ estimates being biased towards the epochs of observation \citep[e.g.][]{Konopacky2016}. The reason for this is that higher eccentricity modes that correspond to this biased region of $T_o$ parameter space are accentuated with flat priors (see $e$, $T_o$ joint posterior in left panel of Figure \ref{fig:Te_corner}). The observable-based prior accounts for this bias by suppressing these regions (right panel of Figure \ref{fig:Te_corner}). For completeness, 1-D marginalized posteriors for all orbital parameters for HR 8799 b, c, d, and e are presented in the Appendix (Figures \ref{fig:pp_b} -- \ref{fig:pp_e}), and the Monte Carlo chains are available upon request. 

\begin{table*}[h!]
\caption{\label{tab:HRinc} Inclination estimates for the HR 8799 planets derived with uniform priors and with the observable-based prior}
\begin{center}
\begin{tabular}{@{}lccc}
\noalign{\hrule height 1pt}
Planet  & $^\mathrm{a}$Prior Ranges (deg) & $^\mathrm{b,d}$Observable-based Prior (deg)    & $^\mathrm{c,d}$Uniform Priors (deg)            \\
\hline
HR 8799b 	 &[0, 180]		&33.4 {\tiny $^{+6.6}_{-4.3}$ }			&35.1 {\tiny $^{+7.0}_{-5.1}$ }				\\	
HR 8799c		 &[0, 180]		&26.7 {\tiny $^{+3.4}_{-11.2}$ }			&26.9 {\tiny $^{+2.7}_{-11.8}$ }				\\	    
HR 8799d		 &[0, 180]		&35.5 {\tiny $^{+5.6}_{-7.8}$ }			&38.1 {\tiny $^{+9.9}_{-13.5}$ }					\\		 
HR 8799e		 &[0, 180]		&25.0 {\tiny $^{+6.2}_{-9.5}$ }			&17.5 {\tiny $^{+9.0}_{-10.7}$ }	 				\\		 
\hline 
\noalign{\hrule height 1pt}
\multicolumn{4}{@{}l}{\noindent \tiny $^\mathrm{a}$ Prior ranges for the observable-based prior are still specified in model-parameter space.}\\
\multicolumn{4}{@{}l}{\noindent \tiny $^\mathrm{b}$ Inclination posterior median with the associated 68\% credible interval, derived using the observable-based prior.}\\
\multicolumn{4}{@{}l}{\noindent \tiny $^\mathrm{c}$ Inclination posterior median with the associated 68\% credible interval, derived using uniform priors.}\\
\multicolumn{4}{@{}l}{\noindent \tiny $^\mathrm{d}$ Monte Carlo chains are made available so that posterior distributions can be evaluated independently since MAP estimates differ slightly from the median.}\\
\end{tabular}
\end{center}
\end{table*}

\subsubsection{Expected Information Gained}\label{sec:EntropyStats}
Biases introduced in the previous analysis affect the posterior information content. Ideally, a prior should be chosen such that the data contribute maximally to the posteriors ({\em e.g.} the prior, as compared to the likelihood, adds the least amount of information possible to the posterior).  For a given dataset $\mathscr{D}$, this is equivalent to maximizing the relative entropy \citep{kullback1951} between the posterior and prior of some model parameter set $\mathscr{M}$,
\begin{equation}\label{eq:dkl}
 \kappa \equiv \int d\mathscr{M}~\mathcal{P}(\mathscr{M} | \mathscr{D})\log \left[\frac{\mathcal{P}(\mathscr{M} | \mathscr{D})}{\mathcal{P}(\mathscr{M})}\right].
\end{equation}
Equation \ref{eq:dkl} is not an ideal measure of information gained as it assumes only one possible dataset.  Instead, the average relative entropy between the posterior and the prior, or the average information gained in the posterior over the prior (defined as ``expected information''  by \citet{Lindley1956}), can be used for this purpose:
\begin{eqnarray}
 \mathcal{I} & \equiv & \int d\mathscr{D} ~\mathcal{P}(\mathscr{D})\int d\mathscr{M} ~\mathcal{P}(\mathscr{M} | \mathscr{D})\log \left[\frac{\mathcal{P}(\mathscr{M} | \mathscr{D})}{\mathcal{P}(\mathscr{M})}\right]\label{eq:exinfoA}\\
 & = & \int \int d\mathscr{D}~ d\mathscr{M} ~\mathcal{P}(\mathscr{M}, \mathscr{D})\log \left[\frac{\mathcal{P}(\mathscr{M}, \mathscr{D})}{\mathcal{P}(\mathscr{D}) \mathcal{P}(\mathscr{M})}\right]\label{eq:exinfoB}\\
 & = & \int d\mathscr{M} ~\mathcal{P}(\mathscr{M})\int d\mathscr{D} ~\mathcal{P}(\mathscr{D} | \mathscr{M})\log \left[\frac{\mathcal{P}(\mathscr{D} | \mathscr{M})}{\mathcal{P}(\mathscr{D})}\right].\label{eq:exinfoC}
\end{eqnarray}
Other objective-prior frameworks \citep[e.g.][]{Berger2009} derive priors by maximizing the expected information. The sampling algorithm that we use to estimate the expected information is described in the Appendix (Algorithm \ref{alg:ei}).

Table \ref{tab:expected_info} lists the expected information -- information gained in the posteriors over the prior -- for orbital fits of each of the HR 8799 planets. When assuming an observable-based prior rather than a uniform prior, the expected information increases by $\backsim$29\% for HR 8799b, $\backsim$26\% for HR 8799c,  $\backsim$35\% for HR 8799d, and $\backsim$33\% for HR 8799e. This increase in expected information indicates that with respect to the prior, the information contained in the data contributes 25 -- 35\% more to the posterior for each of the HR 8799 planets when assuming an observable-based prior.
\begin{table*}[h!]
\caption{\label{tab:expected_info}Expected information gained in the posterior over the prior for the HR 8799 planets.}
\begin{center}
\begin{tabular}{@{}cccc}
\noalign{\hrule height 1pt}
Planet & $^\mathrm{a}$Uniform Prior & $^\mathrm{b}$Observable-based & $^\mathrm{c}$ Percent\T\\
  & (Commonly Assumed) & Prior & Increase\B\\
\hline
HR 8799b & 16.4 $\pm$ 0.1 & 21.2 $\pm$ 0.1 & 29.3  $\pm$ 0.1 \%  \T\\
HR 8799c & 16.7 $\pm$ 0.1 & 21.0 $\pm$ 0.1 & 25.7 $\pm$ 0.1 \% \\
HR 8799d & 16.1 $\pm$ 0.2 & 21.7 $\pm$ 0.3 & 34.8 $\pm$ 0.4 \% \\
HR 8799e & 14.0 $\pm$ 0.1 & 18.6 $\pm$ 0.4 & 32.9  $\pm$ 0.4 \% \B\\
\noalign{\hrule height 1pt}
\multicolumn{4}{@{}l}{\noindent \tiny $^\mathrm{a}$ Average relative entropy between posteriors and uniform priors as a measure of information gained (Sect. \ref{sec:EntropyStats}).}\\
\multicolumn{4}{@{}l}{\noindent \tiny $^\mathrm{b}$ Average relative entropy between posteriors and observable-based priors as a measure of information gained (Sect. \ref{sec:EntropyStats}).}\\
\multicolumn{4}{@{}l}{\noindent \tiny $^\mathrm{c}$ Observable-based priors allow information from data to contribute 25 -- 35\% more to resulting inferences than it could with uniform priors. }\\
\\
\end{tabular}
\end{center}
\end{table*}

\section{Discussion}\label{sec:discussion}
Many stars in the Galactic center and many directly-imaged exoplanets have low orbital phase coverage, causing data to have low constraining power. Prior assumptions dominate parameter estimates in these low-phase-coverage regimes, potentially introducing biases in fitted parameters and producing inaccurate confidence intervals. Uniform priors, commonly assumed in orbit fitting, exacerbate these issues in regions of prior-dominance. In this paper, we propose a new prior that is based on uniformity in observable space rather than in model parameter space to limit the impact of subjective model selection. Statistical tests applied to both simulated and real GC and HR 8799 data indicate that observable-based priors perform better than uniform priors in prior-dominated regimes. 

\subsection{Galactic Center Orbits}\label{sec:GCdiscussion}
Orbits of the short-period stars within the central arcsecond (the S-stars) can be used to probe the dynamics of the GC. There are currently $\sim$40 S-stars with measured orbits, 17 of which have been used in a multi-star fit to constrain the central potential \citep{Gillessen2017}. Although this multi-star fit reduces the uncertainty on fundamental parameters such as the mass of and distance to the central SMBH, it is essential to ensure that accuracy is not jeopardized for this precision. As such, we must ensure that using low-phase-coverage stars in a multi-star fit does not introduce biases due to the statistical effects of prior dominance. In the multi-star fit from \citet{Gillessen2017}, each star is weighted according to the number of data points it contributes to account for biases; however, simulated S0-2 test cases indicate that bias is correlated with orbital phase coverage, though not {\em necessarily} with the number of observations. This is evidenced by the fact that Test Case 1 and Test Case 2 have fewer data points than do all other S0-2 test cases (Table \ref{tab:params}), yet have higher phase coverage and smaller bias values. However, the number of data points or cadence of observations may affect prior performance for orbits with different angular orientations or eccentricities. As such, future work includes testing the performance of the observable-based prior for additional S-stars, and using information on how each star affects global parameter biases to appropriately weight the contribution of each star in a multi-star fit. 
 
In addition, orbital estimates of the S-stars are used to test formation hypotheses that attempt to explain the observed abundance of early-type stars in the GC -- the so-called ``paradox of youth" \citep[e.g.][]{Morris1993, Ghez2003}. For example, the eccentricity distribution of the S-stars can be compared to distributions expected for different formation scenarios \citep[e.g.][]{Gillessen2017, Gillessen2009b, Perets2009, Chen2014, Madigan2014}. One proposed mechanism of S-star formation is tidal capture of a binary star system, which would result in an ejected hyper velocity star and a highly eccentric bound star whose orbit would circularize over a relaxation timescale \citep[e.g.][]{Hills1988, Brown2015}. Whereas a relaxed stellar system expects a thermal distribution, this binary capture scenario expects the eccentricity distribution to peak towards higher eccentricities since the two-body relaxation time ($\sim10^9$ years; \citet{Perets2007}) is an order of magnitude longer than the maximum lifetime of a B star ($\sim10^8$ years). On the other hand, a lower-than-thermal distribution may indicate a disk-migration scenario. \citet{Gillessen2009b} find that the eccentricity distribution is consistent with a thermal distribution to within 3-$\sigma$, though they highlight a slight peak towards higher eccentricities. With a larger sample of stars, \citet{Gillessen2017} later find that the eccentricity distribution indeed is consistent with a thermal distribution, leaving inconclusive evidence of the formation history. We can also look at the orbital plane orientations of the S-stars to see if they are compatible with the clockwise stellar disk. The inclination and angle of ascending node distributions reported by \citet{Gillessen2009b} and confirmed by \citet{Gillessen2017} indicate that a majority of the S-stars have randomly distributed orientations and do not appear to be associated with the clockwise disk of stars located outside the central arcsecond. To confirm these findings or perform more robust tests of these formation scenarios, we must ensure that the measured orbits of the early-type stars with low phase coverage are not biased. Future work includes using observable-based priors to test the effects of the prior on stars with different eccentricity distributions and angular orientations.

\begin{figure}
\begin{center}
\includegraphics[width=\hsize]{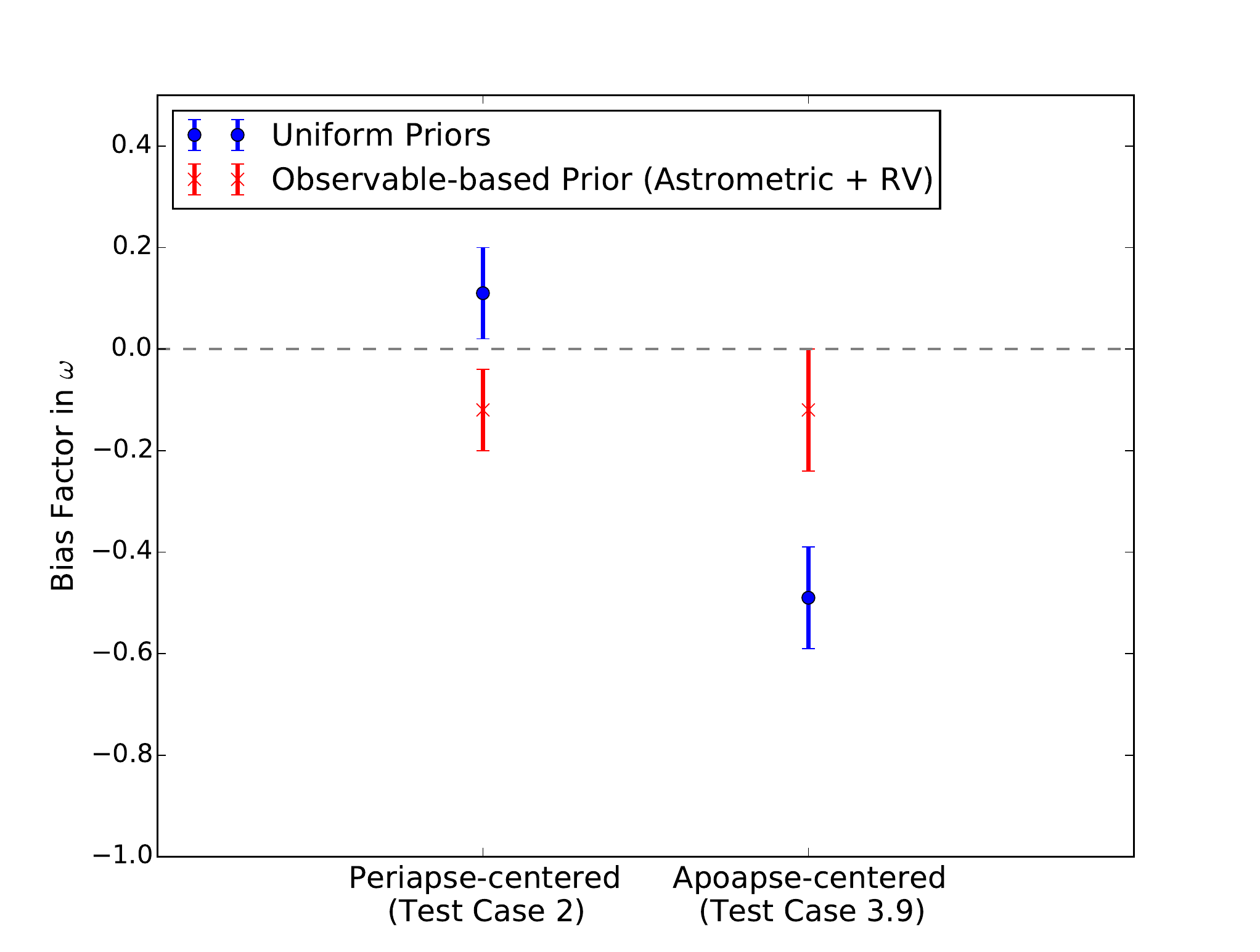}
\caption{\label{fig:dw}  Bias Factor (Section \ref{sec:BiasStats}) for the argument of periapse $\omega$ derived from simulations with uniform priors (blue) and observable-based priors (red). Values are plotted for Test Cases 2 and 3.9 (Table \ref{tab:params}), which are centered on S0-2's periapse and apoapse with $\sim$86\% and $\sim$16\% angular phase coverage, respectively. These two simulations both cover $\sim$50\% of the orbit on the plane of the sky, making it an interesting comparison. Differences in $\omega$ between two such fits could in theory hint at a detection of the precession of the periapse \citep{Parsa2017}; however, the fact that Bias Factor values with the observable-based prior are consistent between the two tests while values with uniform priors differ by over 0.5-$\sigma$ indicates that differences in $\omega$ could in fact be due to statistical biases if prior effects are not considered.}
\end{center}
\end{figure}

Accurate orbital estimates in the GC are also critical as S0-2 has recently gone through its closest approach to the SMBH in 2018, and small deviations from a Keplerian orbit are under investigation. Since S0-2 has full phase coverage, Keplerian orbital estimates with the observable-based prior are consistent with previously published results. However, for a given data set, prior choice becomes more important as the size of the effect under consideration decreases and the complexity of the likelihood increases \citep{Gelman2017}, indicating that prior choice may play a larger role in deciphering small post-Newtonian effects. Future work includes extending this bias analysis to post-Newtonian parameters in General Relativity. Without considering prior influences on posteriors, statistical aberrations may be confused with actual physical processes. In prior-dominated regimes, statistical effects can obscure physical effects such as those from General Relativity \citep[e.g.][]{Hees2017, Parsa2017, Grould2017}. For example, biases induced by prior dominance can cause inferred model parameters to differ when we fit an orbit to the upper and lower portions of S0-2's trajectory independently. Note that because of the angular orientation of S0-2's orbit, the on-sky projection of the orbit in the left panel of Figure \ref{fig:obs_data} does not properly convey how eccentric the true 3D orbit is. As such, angular phase coverage differs greatly from the apparent coverage on the plane of the sky. For example, simulated data in Test Case 3.9 (apoapse-centered) and Test Case 2 (periapse-centered) both cover  $\sim$50\% of the orbit on the plane of the sky, though they differ greatly in angular phase coverage ($\sim$16\% and $\sim$86\% of the 3D orbit based on true anomaly, respectively). Because of this difference in angular phase coverage and consequent difference in information content, the Bias Factor on the argument of periapse $\omega$ differs between these two simulations by over 0.5-$\sigma$ with uniform priors (Figure \ref{fig:dw}). With the observable-based prior, however, the Bias Factor on $\omega$ remains consistent between the two test cases (Figure \ref{fig:dw}), indicating that differences in $\omega$ between these apoapse- and periapse-centered fits could be due to statistical effects of prior dominance rather than a hint of the precession of the periapse.

\subsection{Exoplanet Orbits}\label{Sect:planetsdiscussion}
Similarly, physical models for planet formation depend heavily on orbital estimates. The four directly-observed young giant planets that comprise the HR 8799 system provide an unparalleled opportunity to study the formation and evolution of giant planets \citep[e.g.][]{Marois2010}. Accurate constraints on the planets' orbital plane parameters are essential to understanding the system's dynamical history. Consequently, our physical interpretation can be obscured by the statistical effects of low-phase-coverage data that are not rigorously constraining. 
 
We show that prior choice affects the inference of orbital plane parameters, especially eccentricity and inclination -- key parameters that provide constraints on dynamical models. \citet{Konopacky2016} note that low phase coverage can cause eccentricity estimates to be biased high. By calculating the statistical efficiency and Bias Factor on eccentricity for HR 8799d simulations (based on the 68\% error on the mode of the posterior distribution rather than on central credible intervals due to the multi-modality of resulting posteriors), we confirm that eccentricity estimates indeed are biased high when uniform priors are assumed, but are unbiased with the new observable-based prior. Additionally, the errors on the mode of the posterior distribution are no longer under-estimated with the new prior. This improvement in statistical efficiency and reduction in bias allows greater confidence to be placed in eccentricity estimates inferred with an observable-based prior than in those inferred with uniform priors. In fits to the HR 8799 astrometry, lower eccentricity estimates are favored with the observable-based prior than are estimated with uniform priors, allowing for the possibility of nearly circular orbits -- particularly for planets d and e (Figures \ref{fig:pp_b}, \ref{fig:pp_c}, \ref{fig:pp_d},\ref{fig:pp_e}). In contrast to this result, analyses presented by \citet{Wertz2017} do not support assumptions of circular orbits as they estimate an eccentricity of approximately 0.35 for HR 8799d -- similar to the value we infer with uniform priors. Although \citet{Wertz2017} argue against the circular orbit hypothesis, they note that astrometric biases or underestimation of astrometric errors should not be neglected. Building off of this caveat, we suggest that biases and underestimation of errors in posterior parameter estimates -- not just in initial astrometric measurements -- also cannot be ignored. We therefore suggest that prior considerations be taken into account before ruling out any dynamical models.

The relative inclination of the four planets is another open question whose investigation can provide constraints on dynamical models. Evaluating whether the planets have consistent inclinations requires obtaining accurate estimates of the orbital plane parameters. As such, this question has been widely disputed due to the low constraining power of the data. In this paper, we show that the evidence for consistent inclinations of the four HR 8799 planets is stronger when assuming an observable-based prior than when assuming uniform priors (Figure \ref{fig:eVSi_contour}). Early works suggest that the HR 8799 planets do not have similar inclinations \citep[e.g.][]{Currie2012a, Pueyo2015}; however, using a self-consistent data set, \citet{Konopacky2016} found evidence that the orbital planes of the four planets are consistent within 2-$\sigma$. Building off of this work, using the same self-consistent data set, we provide stronger evidence that the four planets have a consistent inclination of $\sim$30$\mathrm{^o}$ to 1-$\sigma$. Although the $\Omega$ estimates are disjointed, there remains a large overlap in $\Omega$ parameter space with the observable-based prior, indicating that coplanar solutions could still be consistent with the data. While \citet{Wertz2017} similarly show that the HR 8799 planets have consistent inclinations between  $\sim$20$\mathrm{^o}$ and 38$\mathrm{^o}$ with respect to the plane of the sky, they use 3-D dynamical modeling techniques to suggest that the system might not be coplanar, at 2-$\sigma$ significance, due to the disjointed $\Omega$ estimates. Future work includes combining an assessment of the three-dimensional orientations of the orbits \citep{Wertz2017} and of the system dynamics and stability with this prior analysis to further assess the possibility of coplanarity. Here, we simply demonstrate that the prior has a profound effect on parameter estimates and consequently on our physical understanding of the system itself. Prior considerations must be taken into account before confirming or denying physical models when fitting data with such low phase coverage.

\subsection{Statistical Context}\label{sect:statsdiscussion}
Basing an objective prior on experimental design is not an unprecedented idea in Bayesian statistical inference. Reference priors and the Jeffreys prior are objective priors that are defined by the structure of the likelihood. Like reference priors, observable-based priors are dependent on a form of the likelihood, with the goal that the resultant inference is maximally dominated by data \citep{Berger2009}. Observable-based priors differ from these paradigms in that they are not based on the asymptotic nature of an experiment.  This protects against the statistical consequences of asymptotic priors, but also implies that our observable-based prior is weakly-informative rather than truly objective \citep{Gelman2017}. There is some subjectivity added by the choice of the prior in observable space and the wide choice of conversions from higher-dimensional parameter space to lower-dimensional observable space (Equation \ref{eq:obsprdef}). For example, we transform from the observables to $e$ and $P$ for the reasons stated in Section \ref{sec:prior}, though we could in theory have transformed to $e$ and $T_o$ instead. In short, while including relevant information that can influence possible observations, we seek to limit prior influence in regions of prior dominance and thus maximize areas of data dominance.  Because we base our prior analysis on possible data that can be observed (not in the asymptotic limit), we do not achieve the same objectivity as standard objective priors. However, since the  prior's parameterization is determined by the observables, it is less subjective than are uniform priors. Thus, our prior analysis lies in between that of truly objective priors and uniform priors, which are commonly assumed in orbit fitting.  

\section{Conclusion}\label{sec:conclusion}
 Data sets with low orbital phase coverage have low constraining power and thus prior assumptions can bias parameter estimates, produce inaccurate confidence intervals, and profoundly impact inferred posteriors. To improve orbital estimates for objects with low phase coverage -- in particular, stars in the Galaxy's central stellar cluster or directly-imaged exoplanets --  we develop a prior framework that is based on uniformity in observable space rather than in model parameter space. This observable-based prior limits prior influence and allows the data to contribute more heavily to resultant posteriors.

Compared to uniform priors, which are commonly assumed in orbit fitting, the observable-based prior reduces biases in model parameters by up to a factor of two and ensures that credible intervals are not under-estimated for simulated Galactic center data with less than $\sim$40\% phase coverage. Applying the new prior to simulated HR 8799d data shows that the observable-based prior can mitigate the known issue that eccentricity estimates are biased high when data are not rigorously constraining.  

While S0-2 astrometric and RV data have full phase coverage and thus high constraining power, HR 8799 astrometric data have low phase coverage without the additional constraints of RV data. Thus, Sect. \ref{sec:S02results} shows that S0-2's orbital solution derived with the new prior is consistent with that published by \citet{Boehle2016}, while Sect. \ref{sec:HRresults} shows that orbital solutions for the HR 8799 planets are impacted by the observable-based prior, thus influencing our physical interpretation of the system. By limiting prior influence in prior-dominated regions and allowing data to have stronger influence over inferred posteriors, we see stronger evidence for lower eccentricity orbits and for consistent inclinations of the four HR 8799 planets at $\sim$30$\mathrm{^o}$ to within 1-$\sigma$, and do not exclude the possibility of coplanarity.

There are innumerable forms that an observable-based prior can take, though we have only specified two in this work. The objective choice of the prior form can vary and thus should be tested for other models. This framework of prior creation and evaluation can and should be extended to different models and data sets to more accurately estimate orbits of objects with low phase coverage. Such applications include (but are not limited to) Galactic center orbits, directly- imaged planetary systems, and visual binaries. 

\section{Acknowledgments}\label{sec:acknowledgments}
We thank the staff of the Keck Observatory, especially Randy Campbell, Jason Chin, Scott Dahm, Heather Hershey, Carolyn Jordan, Marc Kassis, Jim Lyke, Gary Puniwai, Julie Renaud-Kim, Luca Rizzi, Terry Stickel, Hien Tran, Peter Wizinowich, and former director Taft Armandroff for all their help in obtaining observations. We also thank Dimitrios Psaltis, Eric B. Ford, and David W. Hogg for their feedback and contributions. Support for this work at UCLA was provided by the W. M. Keck Foundation, NSF grant AST-1412615, and the Preston Graduate Fellowship. The W. M. Keck Observatory is operated as a scientific partnership among the California Institute of Technology, the University of California and the National Aeronautics and Space Administration. The Observatory was made possible by the generous financial support of the W. M. Keck Foundation. The authors wish to recognize and acknowledge the very significant cultural role and reverence that the summit of Mauna Kea has always had within the indigenous Hawaiian community. We are most fortunate to have the opportunity to conduct observations from this mountain. The Observatory was made possible by the generous financial support of the W. M. Keck Foundation.

\bibliographystyle{apj}
\bibliography{ms.bib}

\begin{appendix}
\section{Expected Information Sampling}
Calculation of the expected information comes directly from Equations \ref{eq:exinfoB} and \ref{eq:exinfoC}, which imply that the expected information is the expectation of $\log[\mathcal{P}(\mathscr{D} | \mathscr{M})/\mathcal{P}(\mathscr{D})]$ over the joint probablity distribution $\mathcal{P}(\mathscr{D}, \mathscr{M})$, {\em e.g.}
\begin{equation}\label{eq:exinfoApp}
 \mathcal{I} \approx \frac{1}{N}\sum_{i = 0}^N \log\left[\frac{\mathcal{P}(\mathscr{D}_i | \mathscr{M}_i)}{\mathcal{P}(\mathscr{D}_i)}\right], ~~~  \mathscr{D}, \mathscr{M} \sim \mathcal{P}(\mathscr{D}, \mathscr{M})
\end{equation}
over some large $N$.  Equation \ref{eq:exinfoApp} implies a simple algorithm to calculate $\mathcal{I}$:  iteratively sample $\mathscr{D}$ and $\mathscr{M}$ from $\mathcal{P}(\mathscr{D}, \mathscr{M})$ and calculate the average $\log[\mathcal{P}(\mathscr{D} | \mathscr{M})/\mathcal{P}(\mathscr{D})]$.  $\mathcal{P}(\mathscr{D}, \mathscr{M})$ is sampled by first sampling $\mathscr{M}$ from the prior, $\mathcal{P}(\mathscr{M})$, and then drawing a mock dataset, $\mathscr{D}$, from the likelihood, $\mathcal{P}(\mathscr{D} | \mathscr{M})$.  This is summarized in Algorithm \ref{alg:ei}.
\begin{figure}[h!]
\centering
\normalsize
\begin{minipage}{.6\linewidth}
\begin{algorithm} [H]
\label{alg:ei}
 \caption{Expected information sampling algorithm.}
 Draw samples, $\{\mathscr{M}\}$, from $\mathcal{P}(\mathscr{M})$\\
 $n \longleftarrow 0$\\
 \While{$n < N$}
 {
    Draw $\mathscr{M}_{\mathrm{mock}}$ from $\{\mathscr{M}\}$\\
    Draw $\mathscr{D}_{\mathrm{mock}}$ from $\mathcal{P}(\mathscr{D} | \mathscr{M}_{\mathrm{mock}})$\\
    Find evidence, $\mathcal{P}(\mathscr{D}_{\mathrm{mock}})$, of $\mathcal{P}(\mathscr{D}_{\mathrm{mock}} | \mathscr{M}) \mathcal{P}(\mathscr{M})$ \\
    $v_n \longleftarrow \log \left[\frac{\mathcal{P}(\mathscr{D}_{\mathrm{mock}} | \mathscr{M}_{\mathrm{mock}})}{\mathcal{P}(\mathscr{D}_{\mathrm{mock}})}\right]$\\
    $n \longleftarrow n + 1$
 }
 $\mathcal{I} \longleftarrow \frac{1}{N}\sum_i v_i$
\end{algorithm}
\end{minipage}
\end{figure}

\section{HR 8799 Orbital Parameter Posteriors}
\begin{figure*}
\begin{center}
\includegraphics[width=\hsize]{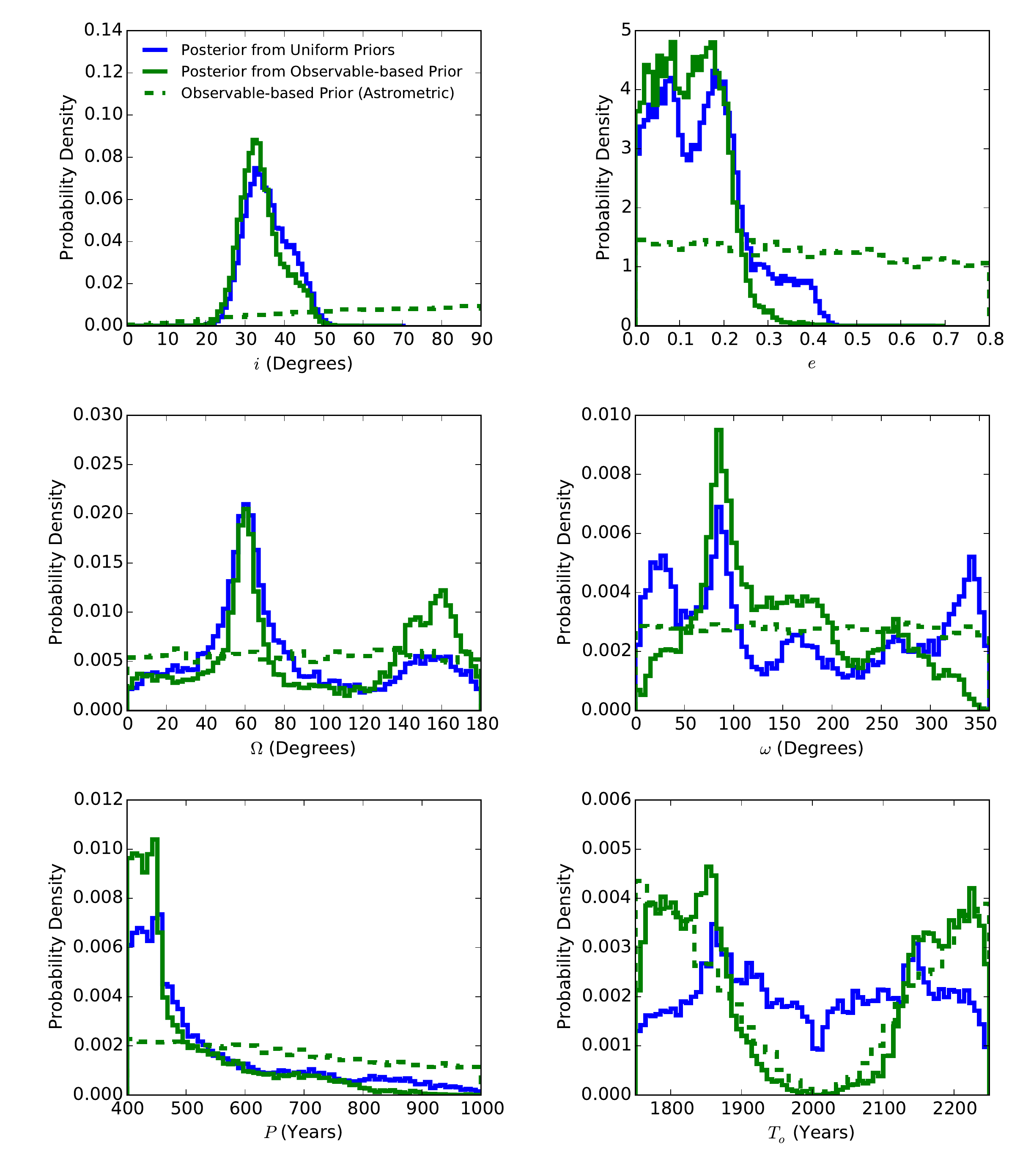}
\caption{\label{fig:pp_b}  Marginalized 1-D posteriors assuming a uniform prior (blue) and a new prior based on uniformity in the astrometric observables (green) for all orbital parameters resulting from a fit to HR 8799b astrometric data from \citet{Konopacky2016}. The observable-based prior itself is also shown with the dashed green line.}
\end{center}
\end{figure*}

\begin{figure*}
\begin{center}
\includegraphics[width=\hsize]{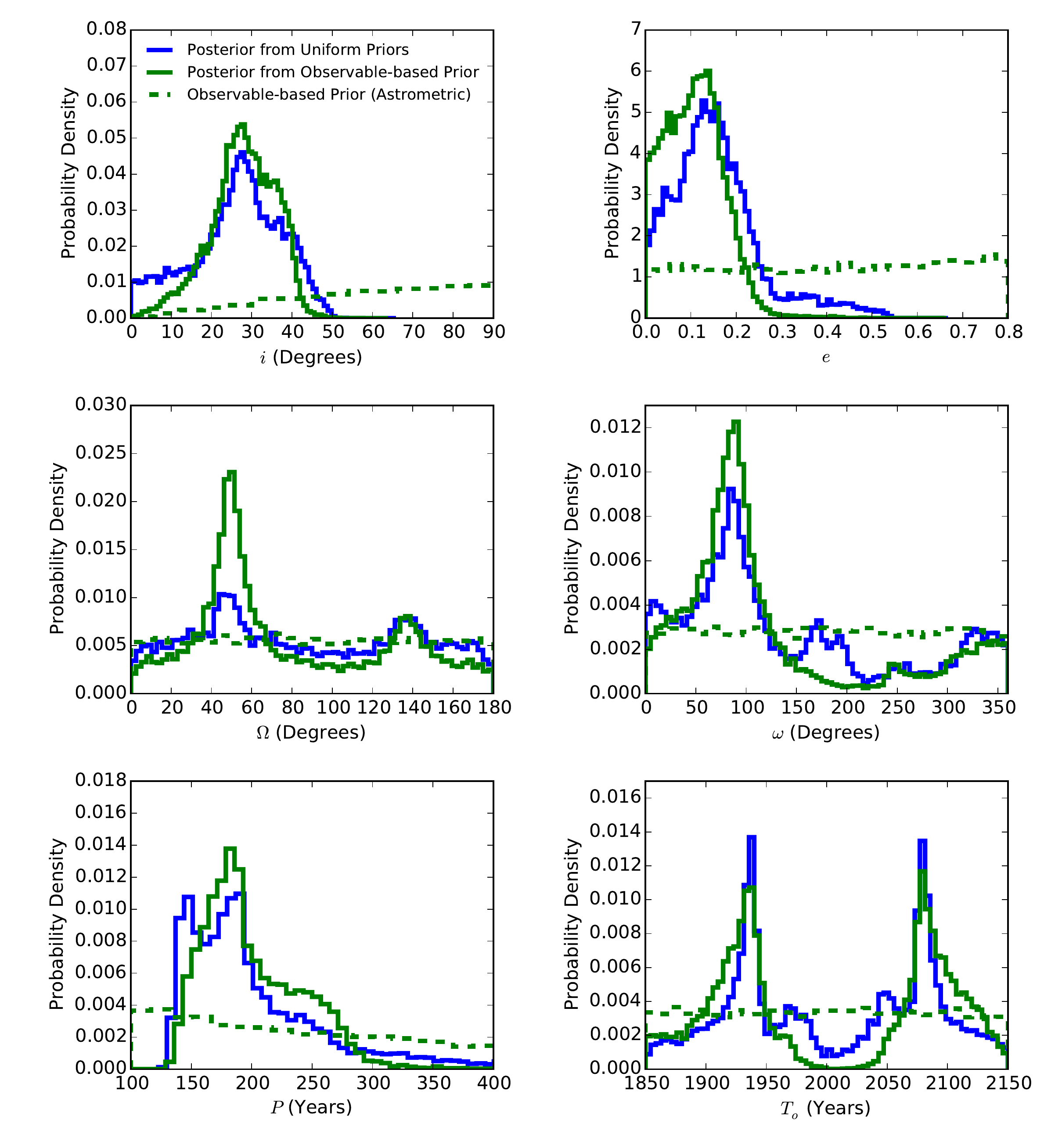}
\caption{\label{fig:pp_c}  Marginalized 1-D posteriors assuming a uniform prior (blue) and a new prior based on uniformity in the astrometric observables (green) for all orbital parameters resulting from a fit to HR 8799c astrometric data from \citet{Konopacky2016}. The observable-based prior itself is also shown with the dashed green line.}
\end{center}
\end{figure*}

\begin{figure*}
\begin{center}
\includegraphics[width=\hsize]{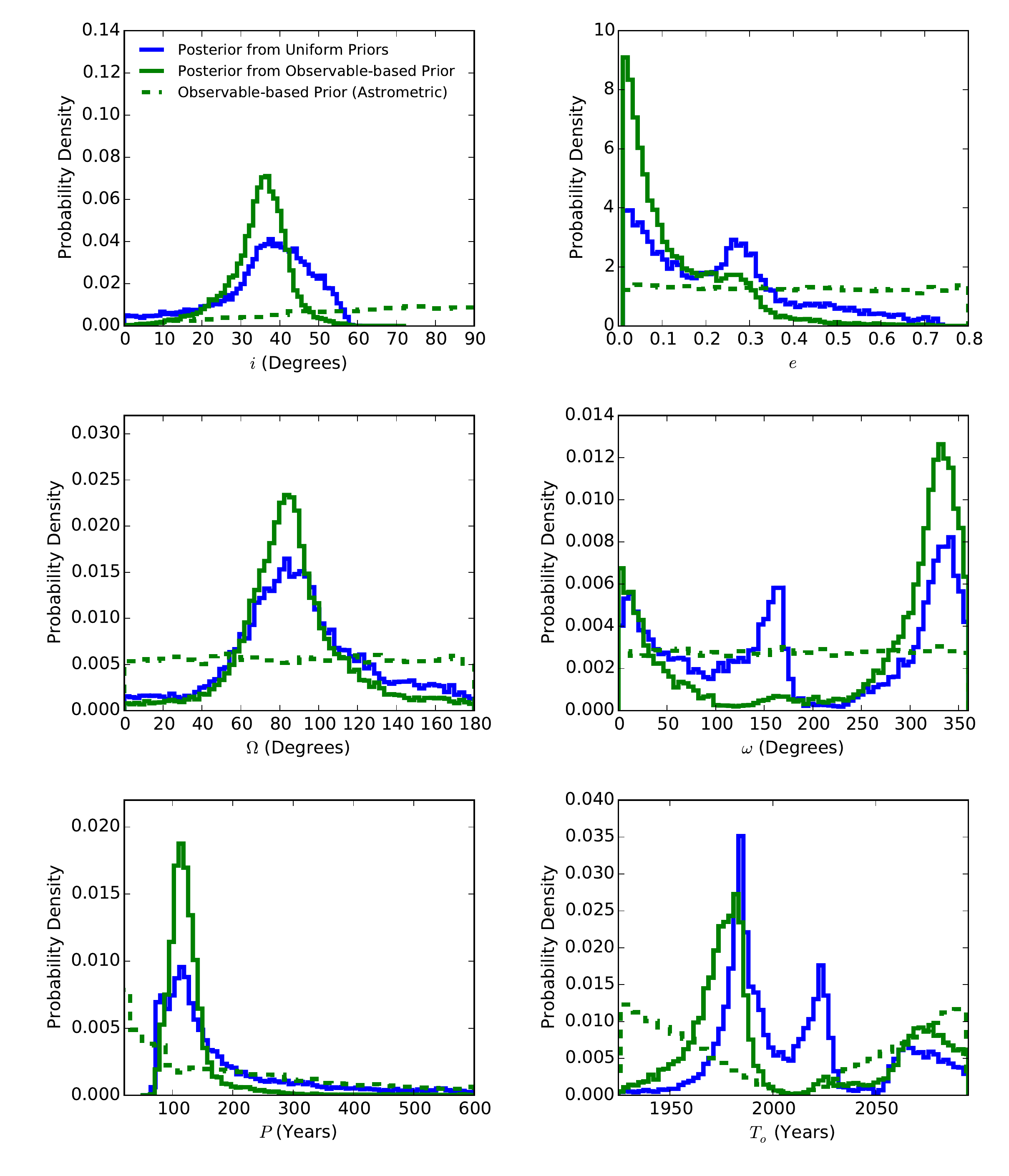}
\caption{\label{fig:pp_d}  Marginalized 1-D posteriors assuming a uniform prior (blue) and a new prior based on uniformity in the astrometric observables (green) for all orbital parameters resulting from a fit to HR 8799d astrometric data from \citet{Konopacky2016}. The observable-based prior itself is also shown with the dashed green line.}
\end{center}
\end{figure*}

\begin{figure*}
\begin{center}
\includegraphics[width=\hsize]{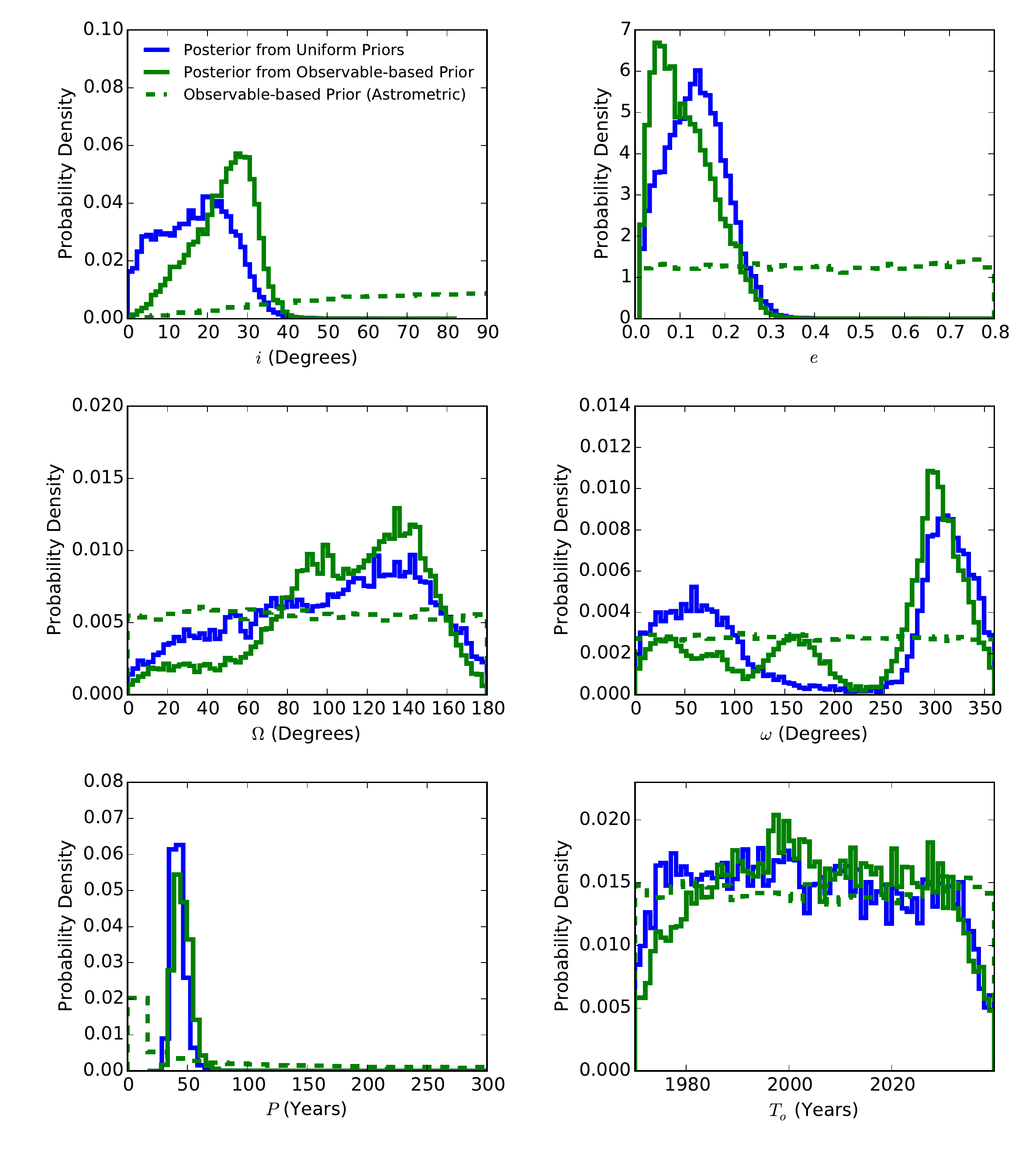}
\caption{\label{fig:pp_e}  Marginalized 1-D posteriors assuming a uniform prior (blue) and a new prior based on uniformity in the astrometric observables (green) for all orbital parameters resulting from a fit to HR 8799e astrometric data from \citet{Konopacky2016}. The observable-based prior itself is also shown with the dashed green line.}
\end{center}
\end{figure*}

Figures \ref{fig:pp_b}, \ref{fig:pp_c}, \ref{fig:pp_d}, and \ref{fig:pp_e} show 1-D marginalized posteriors and the prior for all orbital parameters for HR 8799b, c, d, and e, respectively. Prior choice affects the resulting posteriors, and thus must be taken into consideration. All probability densities, including the prior probability densities, are derived from Monte Carlo simulations. The Monte Carlo chains are also available online. 

Note that the form of the observable-based prior defined in Sect. \ref{sec:prior_orbit} is approximated by summing the Jacobian over all epochs.  For each individual epoch, the prior distribution in $T_o$ should be uniform (modulo boundary effects) according to Equation \ref{eq:Tprior}. However, when summing over a finite number of epochs, the sum is not guaranteed to be flat, as is evident in Figures \ref{fig:pp_b} and \ref{fig:pp_d}.

\end{appendix}
\end{document}